\documentclass[reqno,11pt]{amsart}
\pdfoutput=1

\usepackage{amsmath,amssymb,amsfonts,dsfont,amsthm}    
\usepackage[english]{babel}
\usepackage{graphicx}
\usepackage{float}
\usepackage[font=small,labelfont=bf]{caption}
\usepackage[a4paper]{geometry}
\usepackage{enumerate}
\newtheorem{thm}{Theorem}

\theoremstyle{remark}

\newcommand{\bequ}{\begin{equation}}
\newcommand{\eequ}{\end{equation}}

\usepackage{hyperref}
\hypersetup{pdfborder=0 0 0, 
	    colorlinks=true,
	    citecolor=blue,
	    linkcolor=blue,
	    urlcolor=blue,
	    pdfauthor={Stephane Nonnenmacher}
	   }
\newcommand{\defeq}{\stackrel{\rm{def}}{=}}
\newcommand{\la}{\langle}
\newcommand{\ra}{\rangle}
\newcommand{\eps}{\epsilon}

\newcommand{\cO}{\mathcal O}
\newcommand{\cU}{\mathcal U}

\newcommand{\cD}{\mathcal D}
\newcommand{\cF}{\mathcal F}
\newcommand{\cH}{\mathcal H}
\newcommand{\cL}{\mathcal L}

\newcommand{\cP}{\mathcal P}

\newcommand{\IC}{\mathbb{C}}
\newcommand{\ID}{\mathbb{D}}
\newcommand{\IR}{\mathbb{R}}
\newcommand{\IN}{\mathbb{N}}
\newcommand{\IZ}{\mathbb{Z}}

\newcommand{\Res}{\operatorname{Res}}
\newcommand{\im}{{\rm Im}}
\newcommand{\re}{{\rm Re}}
\def\bbbone{{\mathchoice {1\mskip-4mu {\rm{l}}} {1\mskip-4mu {\rm{l}}}
{ 1\mskip-4.5mu {\rm{l}}} { 1\mskip-5mu {\rm{l}}}}}

\title[Resonances in hyperbolic dynamics]%
{Resonances in hyperbolic dynamics}

\author{St\'ephane Nonnenmacher}
\address{Laboratoire de Math\'ematique d'Orsay, Univ. Paris-Sud, CNRS, Universit\'e
Paris-Saclay, 91405 Orsay, France.}
\email{stephane.nonnenmacher@u-psud.fr}

\begin{document}

\begin{abstract}
The study of wave propagation outside bounded obstacles uncovers the existence of resonances for the Laplace operator, which are complex-valued generalized eigenvalues, relevant to estimate the long time asymptotics of the wave. In order to understand distribution of these resonances at high frequency, we employ semiclassical tools, which leads to considering the classical scattering problem, and in particular the set of trapped trajectories. We focus on ``chaotic" situations, where this set is a hyperbolic repeller, generally with a fractal geometry. In this context, we derive fractal Weyl upper bounds for the resonance counting; we also obtain dynamical criteria ensuring the presence of a resonance gap. We also address situations where the trapped set is a normally hyperbolic submanifold, a case which can help analyzing the long time properties of (classical) Anosov contact flows through semiclassical methods.
\end{abstract}

\maketitle

\section{Introduction}
Spectral geometry attemps to understand the connection between the
{\em shape} (geometry) of a smooth Riemannian manifold $(M,g)$, and
the {\em spectrum} of the positive Laplace-Beltrami operator $-\Delta$
on this manifold. When $M$ is compact, the spectrum is made of
discrete eigenvalues of finite multiplicities $(\lambda_k^2)_{k\geq
  0}$, associated with an orthonormal basis of smooth eigenfunctions $(\phi_k)_{k\geq 0}$. 
What is the role of this spectrum? It allows to explicitly describe the time evolution of the waves waves, e.g. evolved through the wave equation $(\partial^2_{tt}-\Delta)u=0$. 
The connection comes as follows: taking as any initial datum $u(0)=0$, $\partial_t u(0)=u_0\in L^2(M)$, the wave at any time $t\geq 0$ is given by the exact expansion
\bequ\label{e:expansion-compact}
u(t,x) = \big(\frac{\sin(t\sqrt{-\Delta})}{\sqrt{-\Delta}}u_0\big) (x) = \sum_{k\geq 1} \la \phi_k,u_0\ra \,\phi_k(x)\,\frac{\sin(t\lambda_k)}{\lambda_k}\,,\quad x\in M,t\geq 0\,.
\eequ
Hence, any information on the eigenvalues and eigenfunctions allows to better characterize the evolved wave $u(t)$.

\subsection{Scattering}
\begin{figure}[htbp]
\caption{\label{f:scattering}Scattering of a wave by obstacles $\Omega=\cup_i\Omega_i\subset \IR^d$. Parallel lines indicate incoming and outgoing wave trains (arrows indicate the direction of propagation). The blue box indicates a "detector".}
\begin{center} 
\includegraphics[width=.55\textwidth]{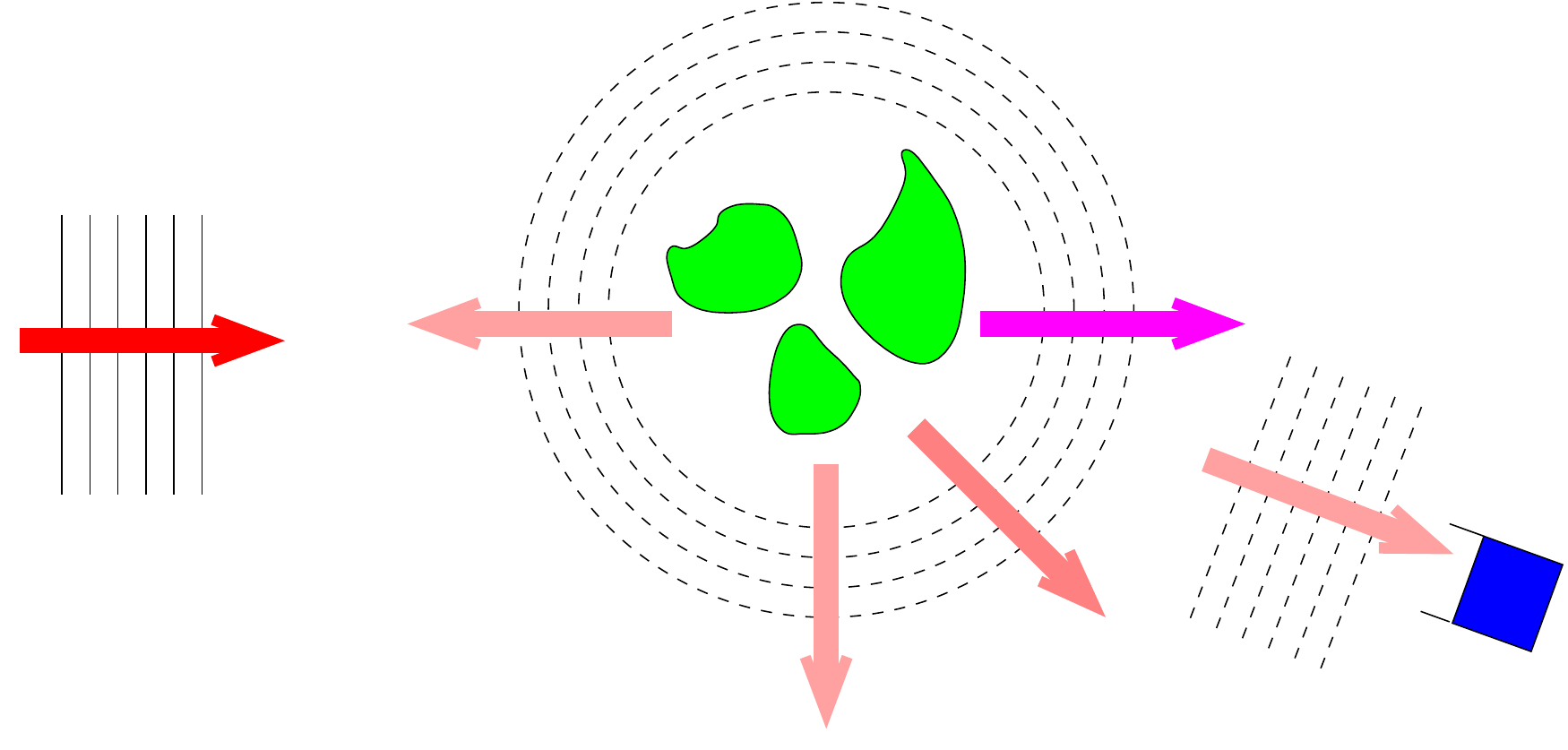}
\end{center}
\end{figure}
In many physical experiments, the waves (or wavefunctions) are not confined to compact domains, but can spread towards spatial infinity. The ambient manifold $(M,g)$ therefore has infinite volume, and in general its geometry towards infinity is "simple". For instance, a physically relevant situation consists of the case where $M=\IR^d\setminus \Omega$, with $\Omega$ an open bounded subset of $\IR^d$, representing a bounded "obstacle" (or a set of several obstacles). These obstacles will {\em scatter} an incoming flux of  waves arriving from a certain direction at infinity, resulting in a flux of outgoing waves propagating towards infinity along all possible directions (see Fig.~\ref{f:scattering}). In actual experiments, the experimentalist can produce incoming waves with definite frequency and direction, and can detect the outgoing waves, along one or several directions. Such an experiment aims at reconstructing the shape of the obstacle, from the analysis of the outgoing waves. 

\subsection{Resonances}
Our objective will not be this ambitious {\em inverse problem}, but we will try characterize quantitatively this scattering phenomenon, assuming some geometric and dynamical properties of the obstacles. This will imply a spectral study of the Laplacian $-\Delta$ on $M$ (say, with Dirichlet boundary conditions on $\partial\Omega$). Due to the infinite volume of $M$, the spectrum of $-\Delta$ is purely continuous on $\IR_+$ with no embedded eigenvalues. However, one can exhibit a form of discrete expansion resembling \eqref{e:expansion-compact} by uncovering {\em resonances} (see e.g. the incoming book \cite{DyZw-book} on scattering and resonances, or the recent comprehensive review \cite{Zw16}).  

Let us assume that the initial datum $u_0\in C^\infty_c(M)$; its time evolution can be expressed through Stone's formula:
\bequ\label{e:Stone}
u(t,x)
= \frac{1}{2i\pi} 
\int_{\IR} d\lambda\, e^{-it\lambda} \,R(\lambda)\,u_0\,,
\eequ
where $R(\lambda)$ is the resolvent operator $(-\Delta-\lambda^2)^{-1}$, first defined in the upper half-plane $\im\lambda>0$, and then continued down to $\lambda\in\IR$  as an operator $L^2_{comp}\to L^2_{loc}$.
$R(\lambda)$ actually admits a meromorphic extension from $\im\lambda>0$ to the full lower half-plane $\IC_-=\{\im\lambda<0\}$ (with a logarithmic singularity at $\lambda=0$ in even dimensions $d$), with the possibility of discrete poles $\{\lambda_k\in \IC_-\}$ of finite multiplicities, called the resonances of the system. 

This meromorphic extension encourages us to deform the contour of the above integral towards a line $C_\gamma=-i\gamma+\IR$, thereby collecting the contributions of the residues at the $\lambda_k$. Assuming that all resonances have multiplicity 1, we obtain the expansion
\bequ\label{e:expansion1}
u(t) = \sum_{\im \lambda_k\geq -\gamma} e^{-it\lambda_k}\,
\Pi_{\lambda_k} u_0 + I(t,C_\gamma)\,,\quad\text{where }\Pi_{\lambda_k} = \frac{1}{2i\pi}\oint_{\lambda_k} R(\lambda)\,d\lambda,
\eequ
\begin{figure}[htbp]
\caption{\label{f:resos-lambda}Contour deformation uncovering resonances of $-\Delta$.}
\begin{center} 
\includegraphics[width=.7\textwidth]{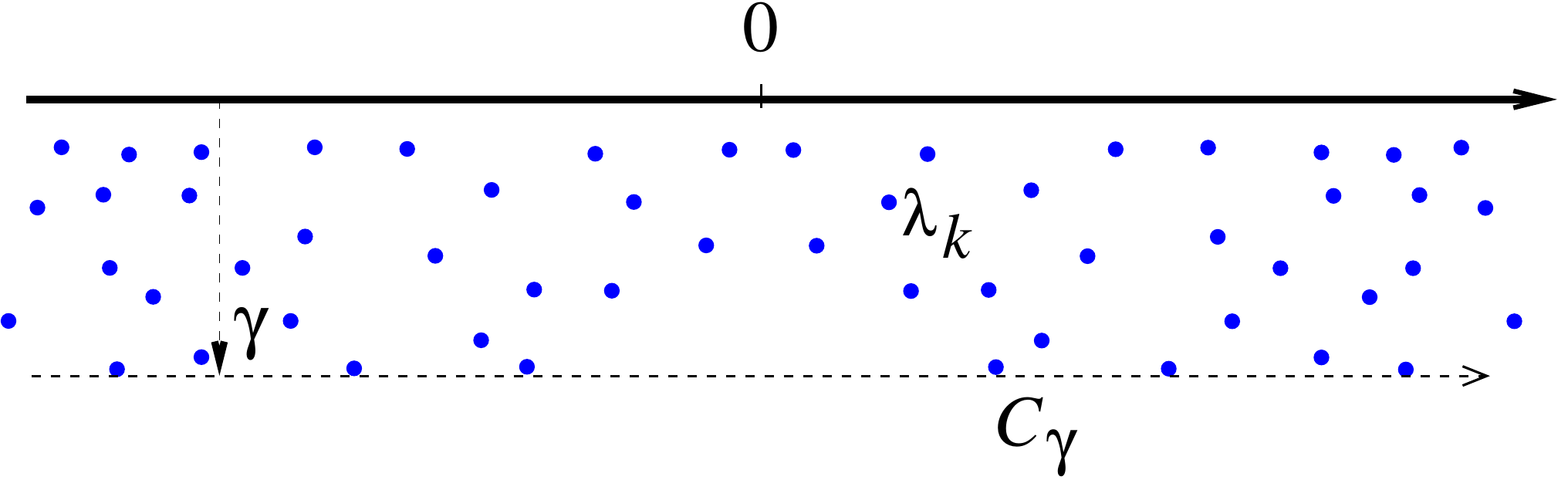}
\end{center}
\end{figure}
and $I(t,C_\gamma)$ is the integral in \eqref{e:Stone} taken along the contour $C_\gamma$. 

Resonances come in symmetric pairs $\lambda_k \leftrightarrow -\bar\lambda_k$ (see Fig.~\ref{f:resos-lambda}). Each $\lambda_k$ corresponds to a resonant state $u_k\in C^\infty_c(M)$, which satisfies the equation $-\Delta u_k = \lambda_k u_k$ and behaves as $\sim e^{i\lambda_k |x|}$ when $|x|\to\infty$, so it diverges exponentially, showing that $u_k\not\in L^2(M)$. If  $\re\lambda_k>0$ the state $u_k$ is said to be {\em purely outgoing}; the complex conjugate function $\bar u_k(x)$ corresponds to the dual resonance $-\bar\lambda_k$ of negative real part: it is purely incoming.
The resonant state $u_k$ allows to express the "spectral projector" $\Pi_{\lambda_k}$ (which acts $L^2_{comp}\to L^2_{loc}$) as 
 $\Pi_{\lambda_k} u_0 = \la \bar u_k, u_0\ra u_k$ (the bracket $\la \bar u_k,u_0\ra=\int dx\,u_k(x)\,u_0(x)$ makes sense since $u_0$ has compact support). 

Assuming we control the size of the remainder term (the contour integral $I(t,C_\gamma)$), the expansion \eqref{e:expansion1} provides informations on the shape and intensity of the wave $u(t,x)$, particularly in the asymptotic $t\gg 1$:  it can explain at which rate the wave leaks ({\em disperses}) out of a given bounded region (say, a large ball $B(R)$), by providing some quantitative bounds on $u(t)\restriction_{B(R)}$. To control the remainder $I(t,C_\gamma)$, one needs to control the size of the truncated resolvent operator $\bbbone_{B(R)}R(\lambda)\bbbone_{B(R)}$ for  $\lambda\in C_\gamma$, in particular the contour should avoid hitting resonances, which requires to control the location of the resonance cloud in the vicinity of $C_\gamma$.

\subsection{Semiclassical regime}\label{s:semiclass}
These arguments hint at our main objective: to determine, as precisely as possible, the distribution of the resonances $\{\lambda_k\}$, and possibly also obtain bounds on the meromorphically continued resolvent $R(\lambda)$. We will be mostly interested in the {\em high frequency regime} $|\re \lambda|\gg 1$, which we choose to rephrase as a {\em semiclassical regime} with small parameter $h\ll 1$. 
To avoid having to deal with both signs of $\re\lambda$, we replace the wave equation by the half-wave equation, written in this semiclassical setting as:
\bequ\label{e:schrod-semiclass}
i h \partial_t u(t) = P_h u(t),\quad\text{with the semiclassical operator }P_h = \sqrt{-h^2\Delta}\,.
\eequ
The small parameter $0<h\ll 1$ is usually called "Planck's constant", since the above equation has the form of a semiclassical Schr\"odinger equation (see below). Here $h$ is just a bookkeeping parameter: we will study the resonances $z_k=z_k(h)\defeq h \lambda_k$ of the operator $P_h$ near some fixed energy $E>0$ (typically $E=1$ for the above half-wave equation), indicating that $\re \lambda_k\sim h^{-1}$. 

We will use the same notations when considering the "true" semiclassical Schr\"odinger equation, describing the evolution of a quantum particle on $M$, subject to an electric potential $V(x)$:
\bequ\label{e:semiclass-schrod}
i h \partial_t u(t) = P_h u(t),\quad P_h = -h^2\Delta + V(x)\,,\quad V\in C^\infty_c(M,\IR)\,.
\eequ
The Schr\"odinger operator $P_h$ also admits resonances $z_k(h)$ in the lower half-plane, obtained as the poles of the resolvent $(P_h-z)^{-1}$,  meromorphically  extended from $\{\re z>0,\im z>0\}$ to $\{\im z<0\}$; now the $z_k(h)$ depend nontrivially of $h$. In these semiclassical notations, the time evolution operator now reads $e^{-itP_h/h}$, so each term $\la \bar u_k,u_0\ra u_k$ in \eqref{e:expansion1} will evolve at a rate $e^{-itz_k/h}$, hence decay at a rate $e^{t\im z_k/h}$. The deeper the resonance ($\equiv$ the larger $|\im z_k|$), the faster this term will decay. We call $\tau_k(h)\defeq \frac{h}{|\im z_k|}$ the lifetime of the resonance. As we will see below, we will be mostly interested in resonances with lifetimes bounded from below, $\tau_k\geq c>0$, which corresponds to studying the resonances in strips of width $\{\im z_k =\cO(h)\}$.
\begin{figure}[htbp]
\caption{\label{f:scattering-class}Left: a wavepacket of wavelength $h$ is scattered by an obstacle. Right:  scattering of classical trajectories (light rays following broken geodesics).}
\begin{center} 
\includegraphics[width=.35\textwidth]{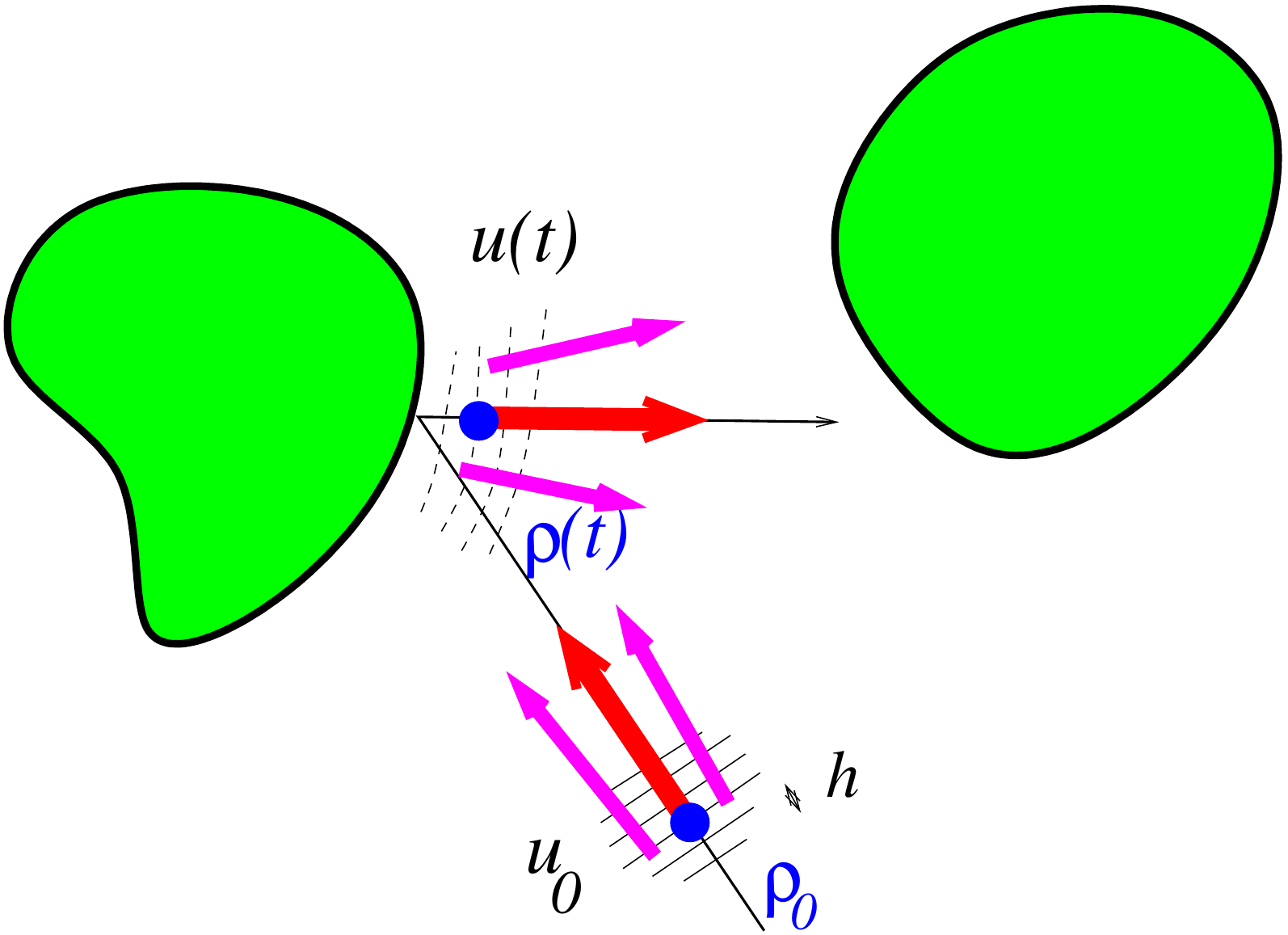}
\hspace{.5cm}
\includegraphics[width=.4\textwidth]{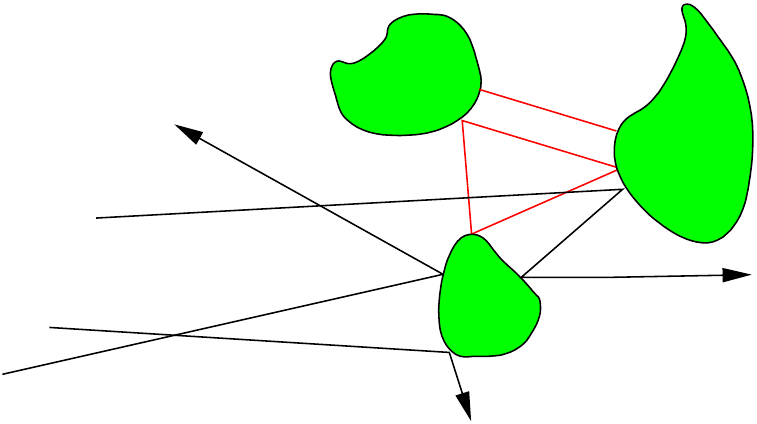}
\end{center}
\end{figure}
\medskip

\subsubsection{Semiclassical evolution of wavepackets}
This semiclassical regime allows us to use the powerful machinery of
semiclassical/microlocal analysis \cite{Zw-book}, which relates the
Schr\"odinger evolution \eqref{e:schrod-semiclass} with the evolution
of classical particles through the Hamiltonian flow $\varphi^t_p$ on
the phase space $T^*M\ni(x,\xi)$. This flow is generated by the
classical Hamiltonian $p(x,\xi)$, given by the principal symbol of the operator $P_h$ (in the above examples $p(x,\xi)=|\xi|$, respectively $p(x,\xi)=|\xi|^2+V(x)$). 
To illustrate this connection, we represent on the left of Fig.~\ref{f:scattering-class} the propagation of a minimum-uncertainty wavepacket $u_0(x)$ through the half-wave equation on $M=\IR^d\setminus\Omega$. The wavepacket can be chosen for instance as a minimum-uncertainty Gaussian wavepacket, also called a {\em coherent state}
$$u_0(x)= C_h\, e^{-\frac{|x-x_0|^2}{2h}}\, e^{i\xi_0\cdot x/h}.
$$
This wavepacket is essentially localized in an $h^{1/2}$-neighbourhood
of the point $x_0$, while its semiclassical Fourier transform
$\cF_hu_0(\xi)$ is localized in an $h^{1/2}$-neighbourhood of the
momentum $\xi_0$ (materialized by the red and pink arrows in the
Figure); we say that this state is  {\em microlocalized} (or centered) on the phase space point $\rho_0=(x_0,\xi_0)$. Heisenberg's uncertainty principle shows that the concentration of such a wavepacket is maximal, equivalently the ``uncertainty" in its position and momentum is minimal.
For a given time window $t\in[0,T]$, in the semicassical limit the evolved state $u(t)=e^{-itP_h/h}u_0$ will remain a microscopic wavepacket, centered at the point $\rho(t)=\varphi^t(\rho_0)$, where $\varphi^t$ is the broken geodesic flow shown on the figure. If we replace the hard obstacles by a smooth potential, the geodesic flow will be replaced by the Hamiltonian flow $\varphi^t_p$. 

\subsubsection{Introducing the trapped set}
In order to analyze the quantum scattering and its associated resonances, it will be crucial to understand the corresponding classical dynamical system, that is the scattering of classical particles induced by obstacles, potentials or metric perturbations, as sketched on the right of Fig.~\ref{f:scattering-class}. In particular, the distribution of resonances will depend on the dynamics of the trajectories remaining in a bounded region of phase space for very long times. For a given energy value $E>0$, we thus introduce the set of points which are trapped forever in the past (resp. in the future, resp. in both time directions):
\bequ\label{e:trapped-set}
\Gamma^\pm_E \defeq \{\rho\in p^{-1}(E),\ \varphi^t_p(\rho)\not\to\infty,\ t\to\mp\infty\},\quad 
K_E = \Gamma^+_E\cap \Gamma^-_E\,.
\eequ
Our assumptions on the structure of $M$ near infinity will always imply that the {\em trapped set} $K_E$ is a compact subset of the energy shell $p^{-1}(E)$; this set is invariant through the flow $\varphi^t_p$. The distribution of the resonances in the semiclassical limit will be impacted by the dynamics of the flow $\varphi^t_p$ on $K_E$. The punchline of the present notes could be:
\begin{quote}
{\em In the semiclassical regime, the distribution of the resonance $\{z_k(h)\}$ near the energy $E$ strongly depends on the structure of the trapped set $K_E$, and of the dynamical properties of the flow $\varphi^t_p$ near $K_E$. }
\end{quote}

\subsection{Hyperbolicity}
In these notes dedicated to ``quantum chaos", we will mostly focus on
systems for which the flow $\varphi^t_p\restriction_{K_E}$ is
hyperbolic (section~\ref{s:NHIM} will contain examples of partial hyperbolicity). What does hyperbolicity mean? It describes the rate at which nearby trajectories depart from each other: for a hyperbolic flow, they separate at an exponential rate, either in the past direction, or in the future, or (most commonly) in both time directions. The trajectories are therefore unstable w.r.t. perturbations of the initial conditions. 
More precisely, an orbit $\cO(\rho_0)=(\varphi^t(\rho_0))_{t\in\IR}\subset p^{-1}(E)$ is hyperbolic if and only if, at each point $\rho\in \cO(\rho_0)$, the $2d-1$-dimensional tangent space $T_\rho p^{-1}(E)$ splits into three subspaces,
$$
T_\rho p^{-1}(E) = \IR X_p(\rho)\oplus E^u(\rho)\oplus E^s(\rho),
$$
where $X_p(\rho)$ is the Hamiltonian vector field generating the flow, $E^s(\rho)$ (resp. $E^u(\rho)$) is the stable (resp. unstable) subspace at $\rho$, characterized by the following contraction properties in the future, resp. in the past:
\bequ\label{e:EuEs}
\exists C,\mu>0, \quad \forall t\geq 0,\quad \|d\varphi^t_p \restriction_{E^s(\rho)}\|\leq C\,e^{-\mu t},\quad  
\|d\varphi^{-t}_p \restriction_{E^u(\rho)}\|\leq C\,e^{-\mu t}\,.
\eequ
\begin{figure}[htbp]
\caption{\label{f:hyperb} Hyperbolicity of the orbit $\cO(\rho)$, with the stable an unstable subspaces transverse to the vector $X_p(\rho)$.}
\begin{center} 
\includegraphics[width=.6\textwidth]{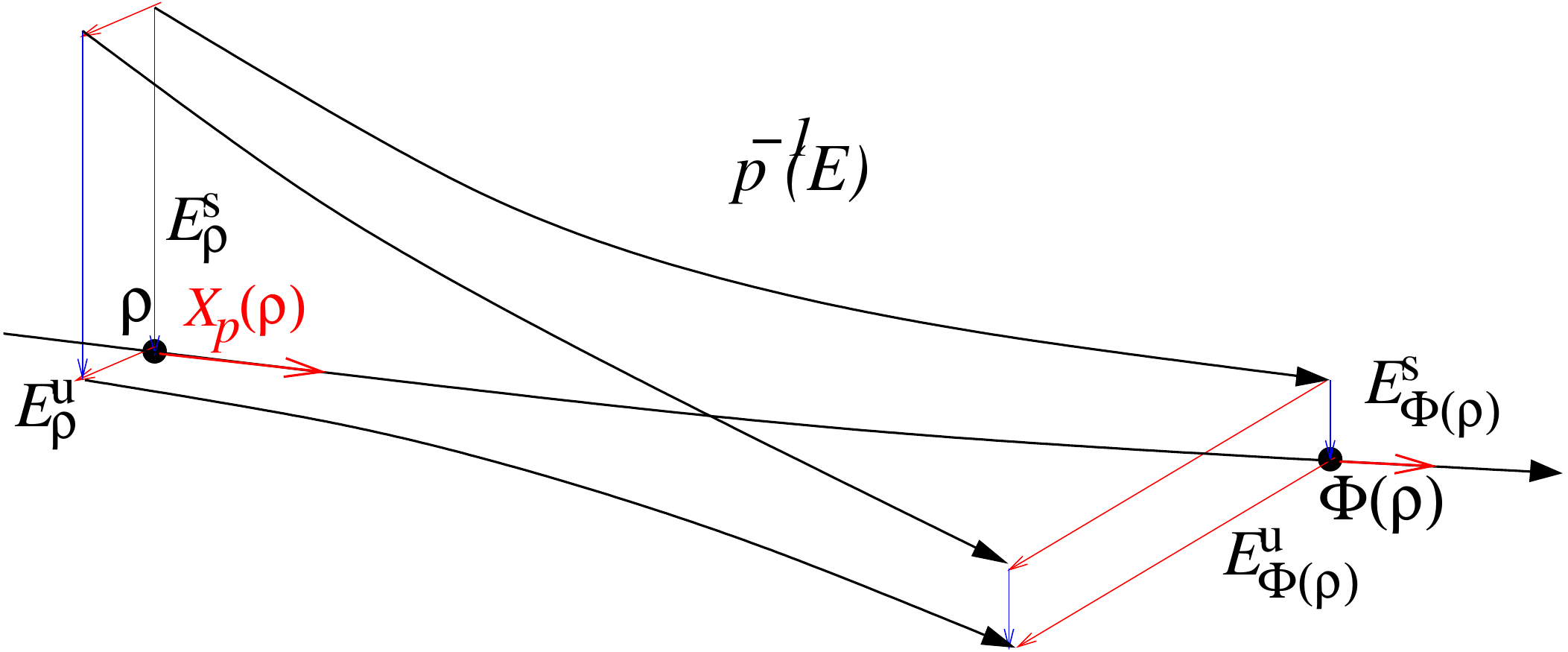}
\end{center}
\end{figure}
The trapped set $K_E$ is said to be (uniformly) hyperbolic if each orbit $\cO(\rho)\subset K_E$ is hyperbolic, with the coefficients $C,\mu$ being uniform w.r.t. $\rho\in K_E$. 
In general the unstable subspaces $E^u_\rho$ are only H\"older-continuous w.r.t. $\rho\in K_E$, even if the flow $\varphi^t$ is smooth; this poor regularity jumps to a smooth (actually, real analytic) dependence in the setting of hyperbolic surfaces described in the next section. 
Such a uniformly hyperbolic flow $\varphi^t_p\restriction_{K_E}$ satisfies Smale's Axiom A; its long time dynamical properties have been studied since the 1960s, using the tools of symbolic dynamics and the thermodynamical formalism \cite{BoRu75}. Below we will use some "thermodynamical" quantities associated to the flow, namely the topological entropy and pressures. The Anosov flows we will mention in the last section are particular examples of such Axiom A flows.

\section{Examples of hyperbolic flows}\label{s:examples}
\subsection{A single hyperbolic periodic orbit}\label{s:single}
\begin{figure}[htbp]
\caption{\label{f:2-obstacles} Left: the simplest case of hyperbolic set: scattering between two strictly convex obstacles. Right: semiclassical resonances for this system}
\begin{center} 
\includegraphics[width=.4\textwidth]{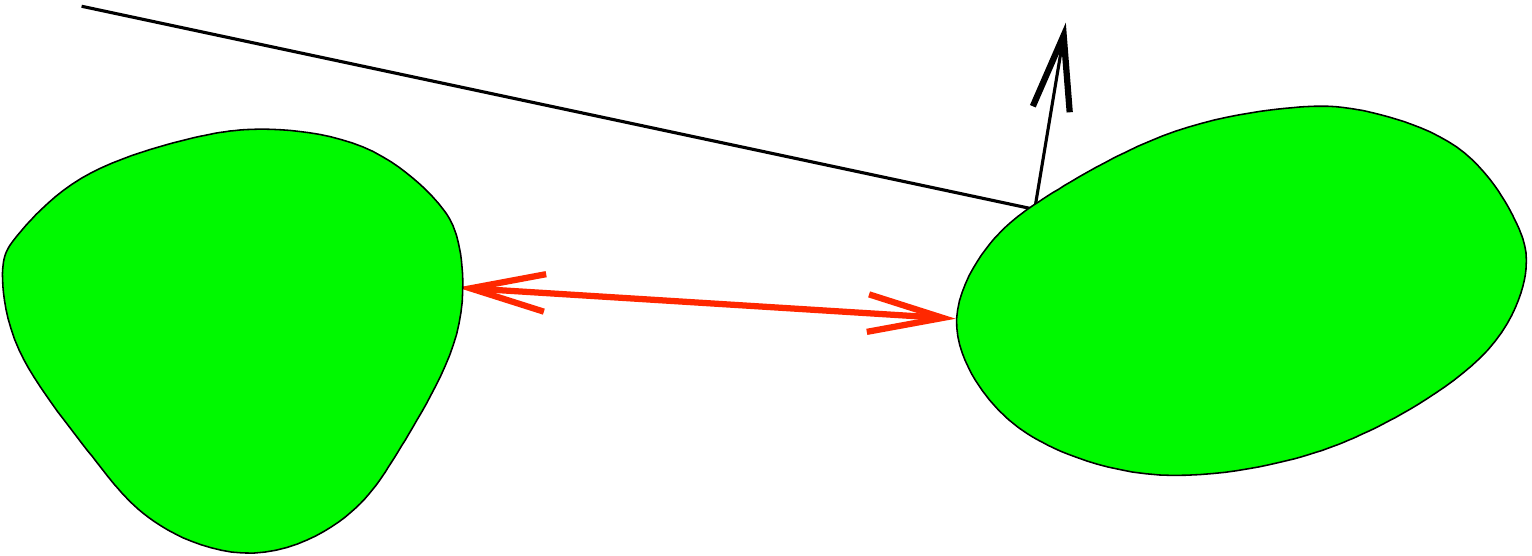}\hspace{.5cm}
\includegraphics[width=.5\textwidth]{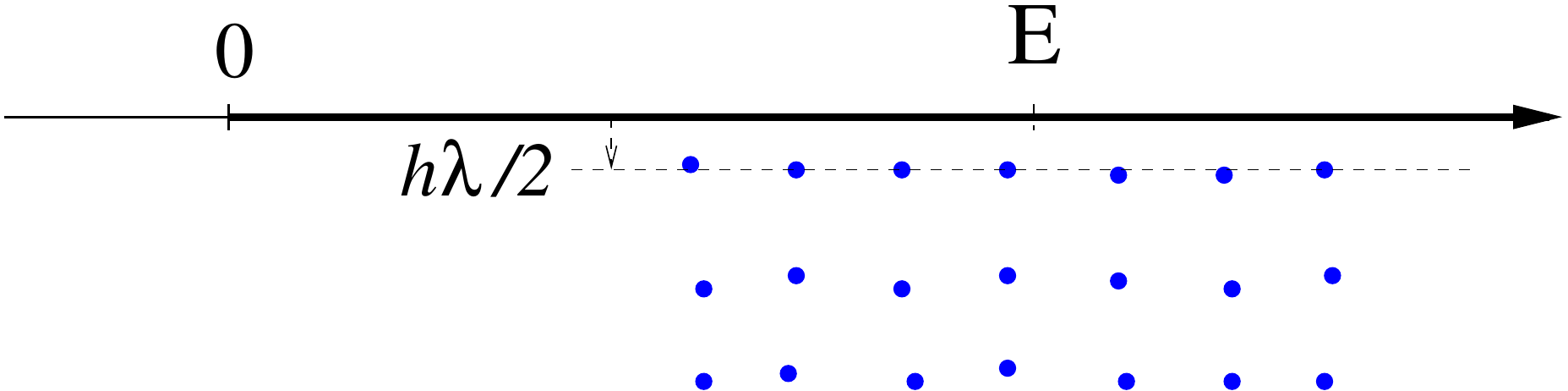}
\end{center}
\end{figure}

The simplest example of hyperbolic set occurs in the scattering by the union of two disjoint strictly convex obstacles in $\IR^d$: in that case the trapped set is made of a single orbit bouncing  periodically between the two obstacles (see Fig.~\ref{f:2-obstacles}). For this simple situation, the resonances of $P_h=-h^2\Delta$ can be computed very precisely in the semiclassical limit \cite{Ika83,GeSj87}; in dimension $d=2$, in a small neighbourhood of the classical energy $E=1$, they asymptotically form a half-lattice:
\bequ\label{e:resos-latt}
z_{\ell,k}(h)= E(h) + \frac{2\pi hk}{T}-ih\lambda(1/2+\ell)+\cO(h^2),\quad \ell\in\IN, \ k\in\IZ,\ E(h)=1+\cO(h)\,.
\eequ
Here $T$ is the period of the bouncing orbit, while $\lambda>0$ is the rate of unstability along the orbit, meaning that 
$\|d\varphi^T_p\restriction_{E^u(\rho)}\| = e^{\lambda T}$. Obtaining such explicit formulas for the resonances is specific to this very simple situation, but it already presents two interesting features. First, the number of resonances in any rectangle $R(E,Ch,\gamma h)$ of the type \eqref{e:rectangle} is uniformly bounded when $h\to 0$, and it is nonzero if $\gamma$ and $C$ are large enough. Second, if $\gamma<\lambda/2$ (and if $h$ is small enough), the box $R(E,Ch,\gamma h)$ will be empty of resonances: this is the first instance of a {\em resonance  gap} connected with the hyperbolicity of the flow on the trapped set.

\subsection{Fully developed chaos: fractal hyperbolic trapped set}\label{s:chaos}
\begin{figure}[htbp]
\caption{\label{f:3-obst} Left: three convex obstacles on $\IR^2$, leading to a fractal hyperbolic repeller. Right: intersection of $K_E$ with a Poincar\'e section $\Sigma$.}
\begin{center} 
\includegraphics[width=.38\textwidth]{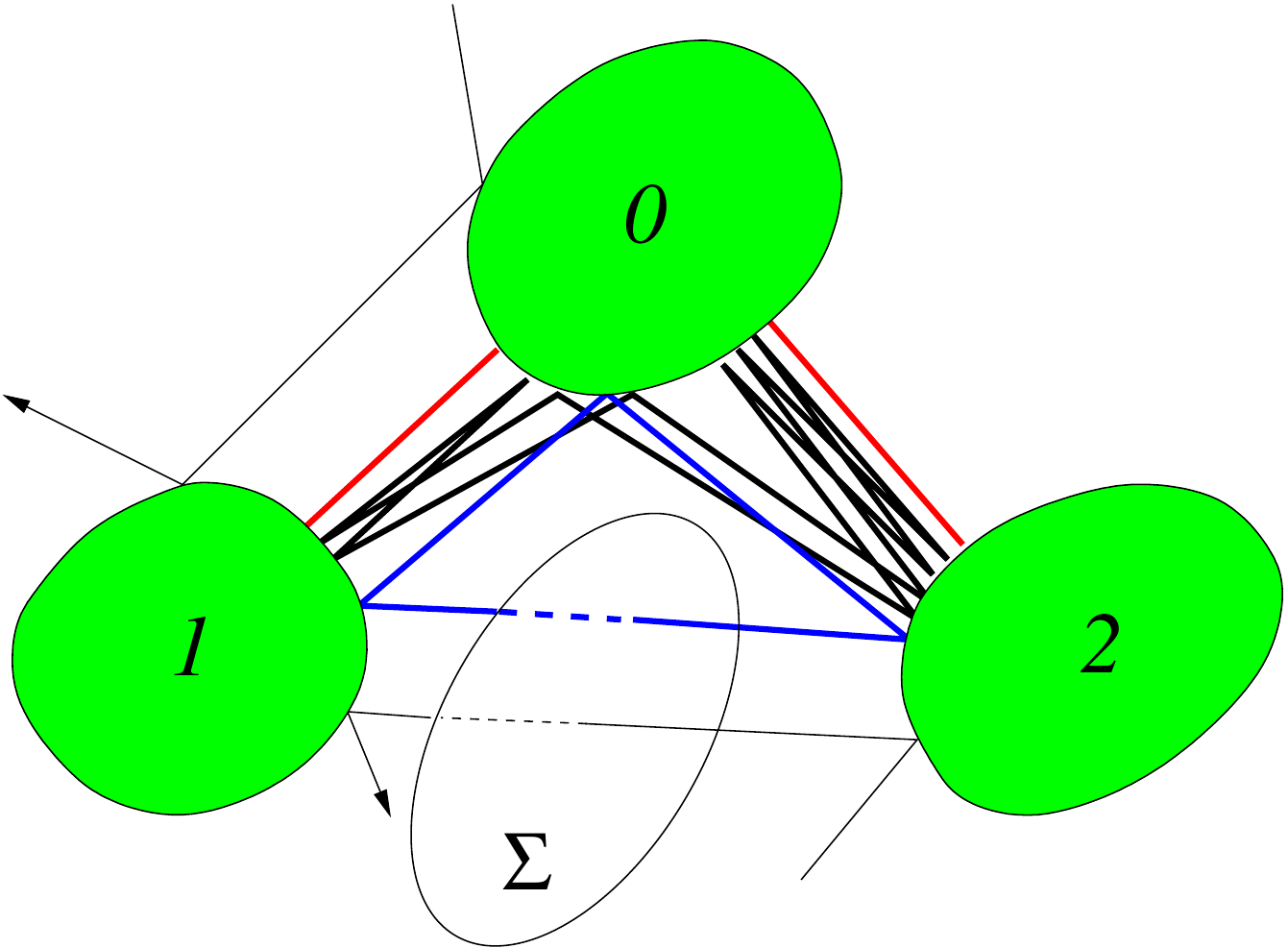}\hspace{2cm}
\includegraphics[width=.26\textwidth]{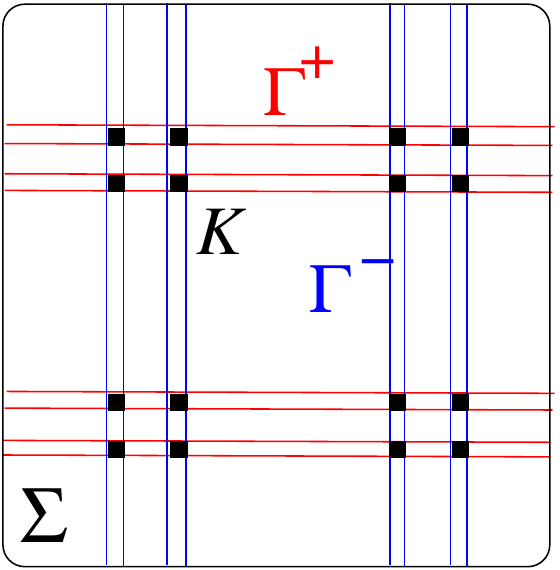}
\end{center}
\end{figure}

Beside hyperbolicity, the second ingredient of ``chaos" is the complexity of the flow, which can be characterized by a positive topological entropy, indicating an exponential proliferation of long periodic orbits: 
\bequ\label{e:entropy}
H_{top}(\varphi^t\restriction_{K_1}) = \lim_{T\to\infty} \frac{1}{T}\log \#\{\gamma\in \text{Per}(K_1),\ T\leq T_\gamma\leq T+1\}\,,
\eequ
where ${\rm Per}(K_1)$ denotes the set of periodic orbits in $K_1$,
and $T_\gamma$ is the period of the orbit $\gamma$. 
A simple example of system featuring such a chaotic trapped set is obtained by adding one convex obstacle to the 2-obstacle example of the previous paragraph. Provided this third obstacle is well-placed with respect to the other two  (so that the three obstacles satisfy a ``no-eclipse condition", like in Fig.~\ref{f:3-obst}, left), the trapped trajectories at energy $E=1$ form a hyperbolic set $K_1$, which contains a countable number of periodic orbits, and uncountably many nonperiodic ones. 
A way to account for this complexity is to construct a symbolic representation of the orbits. Label each obstacle by a number $\alpha\in \{0,1,2\}$; then to each bi-infinite word $\cdots \alpha_{-1}\alpha_0\alpha_{1}\alpha_{2}\cdots$ such that $\alpha_i\neq \alpha_{i+1}$, corresponds a unique trapped orbit in $K_1$, which hits the obstacles sequentially in the order indicated by the word. Periodic words correspond to periodic orbits, nonperiodic words to nonperiodic orbits. This correspondence between words and orbits allows to quantitatively estimate the complexity of the flow on $K_1$. In turn, the strict convexity of the obstacles ensures that all trapped orbits are hyperbolic, the instability arising at the bounces.

The trapped set $K_1$ has a fractal geometry, which can be described by some fractal dimension. It is foliated by the trajectories (which accounts for one "smooth" dimension), so its fractal nature occurs in the transverse directions to the flow, visible in its intersection with a Poincar\'e section $\Sigma\subset S^*X$ (see Fig.~\ref{f:3-obst}). This intersection $K_E\cap \Sigma$ (represented by the union of black squares) has the  structure of a horseshoe; as the intersection of stable ($\Gamma^-$) and unstable ($\Gamma^+$) manifolds, it  locally has a ``product structure".

In space dimension $d=2$, the dimension of $K_1$ can be expressed by using a {\em topological pressure}. This pressure, a ``thermodynamical" quantity of the flow, is defined in terms of the {\em unstable Jacobian} of the flow, $J^u_t(\rho)=|\det(d\varphi^t\restriction_{E^u(\rho)})|$. For a periodic orbit $\gamma$ of period $T_\gamma$, we denote  $J^u(\gamma)=J^u_{T_\gamma}(\rho_\gamma)$, where $\rho_\gamma $ is any point in $\gamma$. Now, for any $s\in\IR$, we may define the pressure as
\bequ\label{e:pressure}
\cP(s)=\cP(s,\varphi^t\restriction_{K_1})=\lim_{T\to\infty} \frac{1}{T} \log \sum_{T\leq T_\gamma \leq T+1} J^u(\gamma)^{-s}\,,
\eequ
where the sum runs over all periodic orbits $\gamma\in {\rm Per}(K_1)$
of periods in the interval $[T,T+1]$. $\cP(0)$ is equal to the
topological entropy \eqref{e:entropy}, which is positive. When increasing $s$, the factors $J^u(\gamma)^{-s}$ decay exponentially when $T\to\infty$, hence the hyperbolicity embodied by these factors balances the complexity characterized by the large number of orbits. The pressure $\cP(s)$ is smooth and strictly decreasing with $s$, and one can show that $\cP(1)<0$; hence, it vanishes at a single value $\delta\in(0,1)$. In the 2-dimensional setting (for which $E^{u/s}(\rho)$ are 1-dimensional), the Hausdorff dimension of $K_1$ is given by Bowen's formula:
$$
\dim K_1 = 1+2\delta \Longleftrightarrow \cP(\delta)=0\,.
$$
The topological pressure will pop up again when studying resonance gaps, see Thm~\ref{th2}.

\subsection{An interesting class of examples: hyperbolic surfaces of infinite area}\label{s:hyperb-surf}
We have mentioned above that one way to "scatter" a wave, or a classical particle, was to modify the metric on $M$ in some compact neighbourhood. Because we are interested in hyperbolic dynamics, an obvious way to generate hyperbolicity is to consider metrics $g$ of negative sectional curvature (giving $M$ locally the surface the aspect of a ``saddle"). Such a metric automatically induces the hyperbolicity of the orbits, the instability rate being proportional to the square-root of the curvature.
\begin{figure}[htbp]
\caption{\label{f:hyperb2} Construction of a hyperbolic surface $M=\Gamma\setminus\IC$ of infinite volume. Left: fundamental domains of the action of $\Gamma$ on $\ID$. Right: representation of $M$.}
\begin{center} 
\includegraphics[width=.7\textwidth]{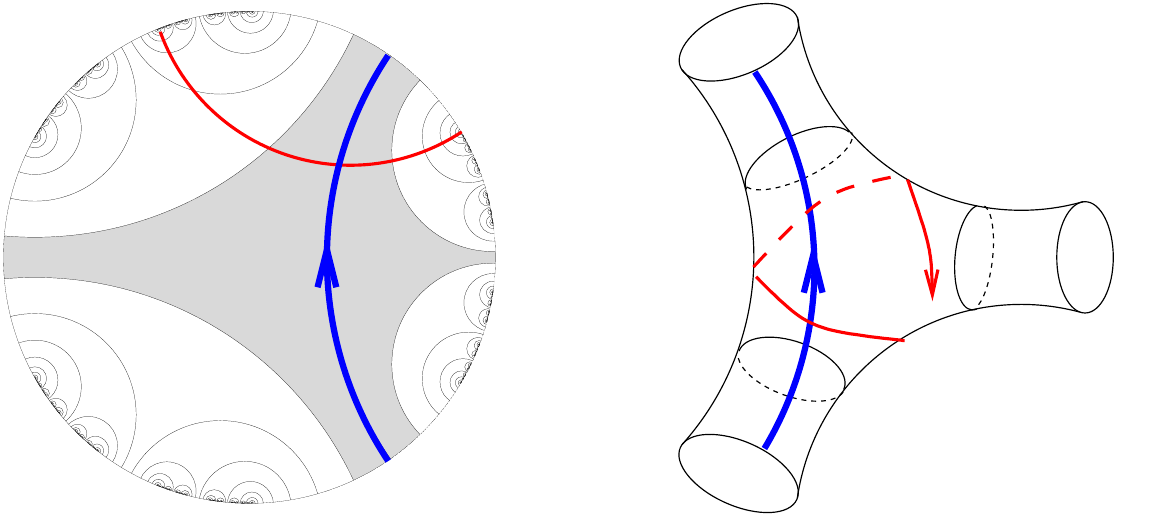}
\end{center}
\end{figure}

Such surfaces can be constructed \cite{Bor16} by starting from the Poincar\'e hyperbolic disk $\ID=\{z\in\IC,\ |z|<1\}$, equipped with the metric $g=4\frac{dz\,d\bar{z}}{(1-z\bar z)^2}$: the curvature is then equal to $-1$ everywhere. The Lie group $SL(2,\IR)$ acts on this disk isometrically. By choosing a discrete subgroup $\Gamma< SL(2,\IR)$ of the Schottky type, the quotient $M=\Gamma\setminus \ID$ is a smooth surface of infinite volume, without cusps. On the left of Fig.~\ref{f:3-obst} we represent the Poincar\'e disk, tiled by fundamental domains of such a Schottky subgroup $\Gamma$ (the grey area is one fundamental domain), the boundaries of the domains being given by geodesics on $\ID$ (which corresond to Euclidean circles hitting $\partial\ID$ orthogonally). On the figure we also notice the accumulation of small circles towards a subset $\Lambda_\Gamma\subset \partial\ID$, called the limit set of the group $\Gamma$. This limit set is a fractal set of dimension $\delta=\delta_\Gamma\in (0,1)$. 

On the right of the figure we plot the quotient surface $M=\Gamma\setminus \ID$, composed of a compact part (the "core") and of three ``hyperbolic funnels" leading to infinity. 
The trapped geodesics of $M$ are fully contained in the compact core, they can be represented by geodesics on $\ID$ connecting two points of $\Lambda_\Gamma$ (red geodesic on the figure). On the opposite, geodesics on $\ID$ crossing $\partial\ID\setminus \Lambda_\Gamma$ correspond to transient geodesics on $M$ (blue geodesic on the figure) which start and end in a funnel. The trapped set can therefore be identified as $K_1\equiv \Lambda_\Gamma\times\Lambda_\Gamma\times\IR$, and its Hausdorff dimension $\dim K_1=1+2\delta$. 

The Laplace-Beltrami operator $-\Delta_M$ has a continuous spectrum on $[1/4,\infty)$, which is usually represented by the values $s(1-s)$, for a spectral parameter $s\in \frac12+i\IR$. The resolvent operator $R(s)=(-\Delta_M-s(1-s))^{-1}$ can be meromorphically extended from $\{\re s>1/2\}$ to $\{\re s<1/2\}$. The resonances are given by a discrete set $\{s_k\}$ in the half-space $\{\re s<1/2\}$. 
A huge advantage of this model, is that these resonances are given by the zeros of the Selberg zeta function
$$
Z_\Gamma(s) \defeq \prod_{\gamma\in\text{Per}^*}\prod_{m= 0}^\infty (1-e^{-(s+m)|\gamma|})\,,
$$
where $\text{Per}^*$ denotes the set of primitive periodic geodesics on $M$. This exact connection between {\em geometric data} (lengths of the periodic geodesics) and {\em spectral data} (resonances of $-\Delta$) is specific to the case of surfaces of constant curvature. Another particular feature of the constant curvature is the fact that the stable/unstable directions $E^{s/u}(\rho)$ can be defined at any point $\rho\in M$, and depend smoothly on the base point $\rho$.  

The identification of the resonances with the zeros of $Z_\Gamma(s)$ provides powerful techniques to study their distribution, with purely "classical" techniques, without any use of PDE methods. These zeros can be obtained by studying a 1-dimensional map on the circle, called the Bowen-Series map, constructed from the generators of the group $\Gamma$ \cite{Nau05}. This map induces a family of transfer operators $\cL_s$ indexed by the spectral parameter; one shows that these operators, when acting on appropriate spaces of analytic functions, are nuclear (in the sense of Grothendieck), and that the Selberg zeta function can be obtained as the Fredholm determinant $Z_\Gamma(s) =\det(1-\cL_s)$. The spectral study of the classical transfer operators $\cL_s$ can therefore deliver informations on the resonance spectrum, which are often more precise than what is achievable through PDE techniques.


\section{Fractal Weyl upper bounds}

\subsection{Counting long living resonances}

\begin{figure}[htbp]
\caption{\label{f:resos33}Resonances of a semiclassical operator $P_h$ in the rectangle $R(E,Ch,\gamma h)$. Right: spectrum of the twisted operator $P_{h,\theta}$.}
\begin{center} 
\includegraphics[width=.4\textwidth]{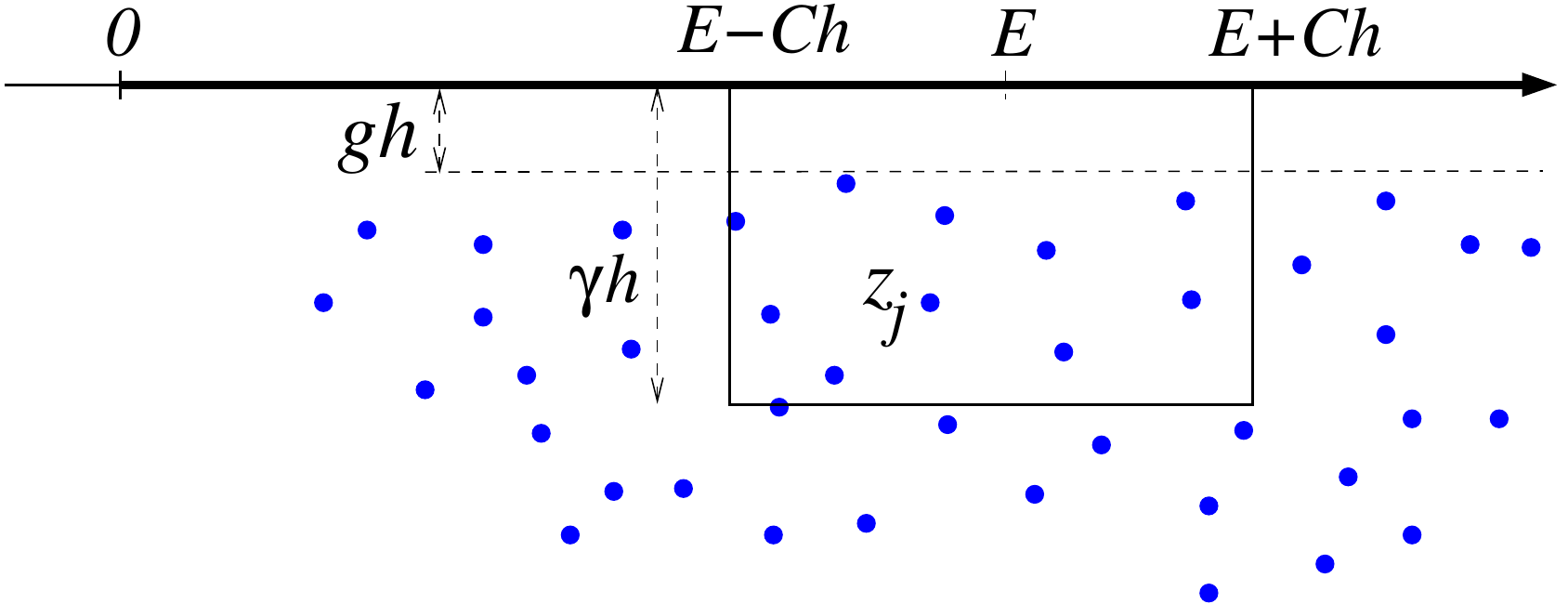}\hspace{.5cm}
\includegraphics[width=.4\textwidth]{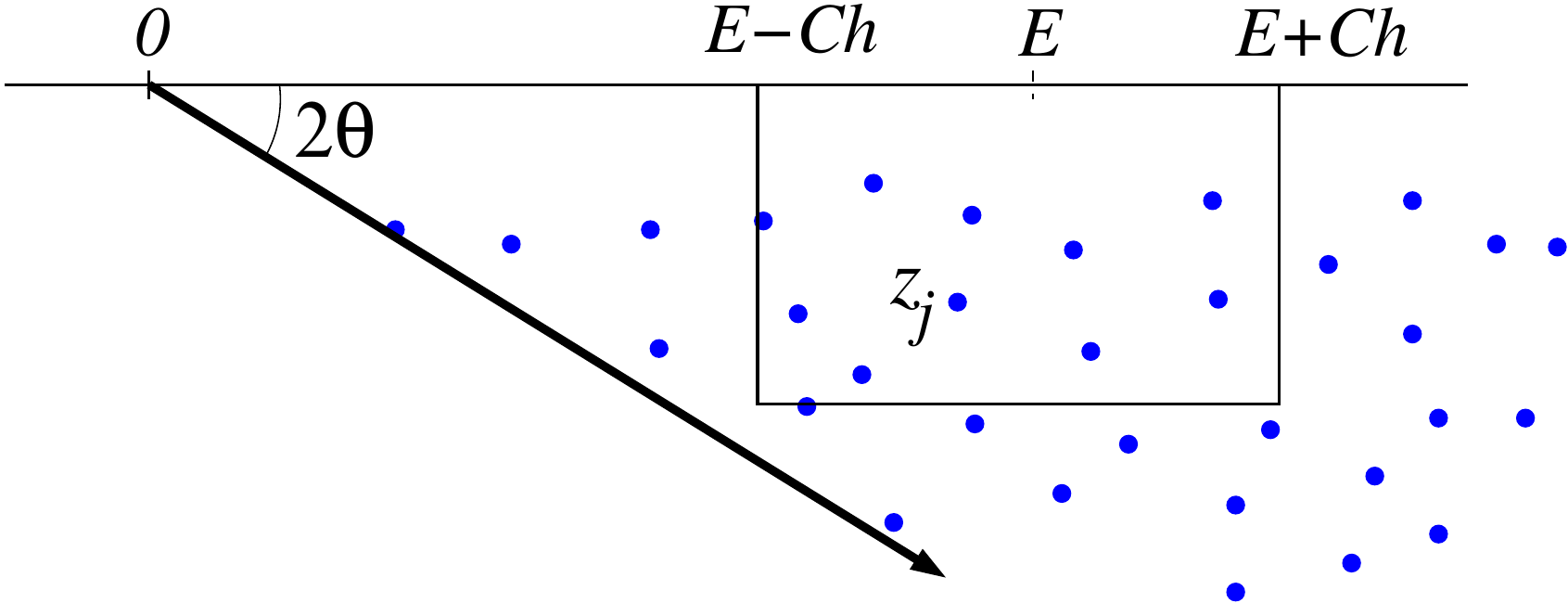}
\end{center}
\end{figure}

We are interested in the distribution of the resonances $(\lambda_j)$ (for $-\Delta$) or $(z_k(h))$ (for $P_h$) in the lower half-plane. Because we want to use these resonances in dynamics estimates as in \eqref{e:expansion1}, we will focus on the {\em long living resonances}, such that $\im z_k(h)\geq - \gamma h$ for some fixed $\gamma>0$, or equivalently such that the corresponding lifetimes $\tau_k(h)\geq 1/\gamma>0$. We will also focus on resonances such that $\re z_k$ lies in some small energy window  $[E-\eps,E+\eps]$: this will allow us to connect their distribution with the properties of the classical flow at energy $E$.
Fig.~\ref{f:resos33} sketches the more precise spectral region we will study, centered at $E>0$: we will count the resonances in rectangles of the type 
\bequ\label{e:rectangle}
R(E,Ch,\gamma h)=[E-Ch,E+Ch]-i[0,\gamma h],\quad C,\gamma>0\ \text{independent of $h$}.
\eequ
In the present section, our main result is a {\em fractal Weyl upper bound} (see Thm~\ref{th1}) for the number of resonances in those rectangles.
In the next section we will be especially interested in situations for which such a rectangle contains {\em no} resonance, like in the rectangle $R(E,Ch,g h)$ of Fig.~\ref{f:resos33}: we will then speak of a (semiclassical) {\em resonance gap} near the energy $E$.

\subsection{Complex deformation of $P_h$: turning resonances into eigenvalues}\label{s:complex-deform}
For simplicity we consider manifolds $M$ which, outside some big ball $B(R_0/2)$, is equal to the Euclidean space $\IR^d\setminus B(R_0/2)$. To analyze the resonances of $P_h$ in $R(E,Ch,\gamma h)$, a convenient method consists in twisting the selfadjoint operator $P_h$ into a nonselfadjoint operator $P_{h,\theta}$, through a ``complex deformation'' procedure \cite{AC71}. Outside a large ball $B(R_0)$, the differential operator $P_{h,\theta}$ is equal to $-h^2e^{-2i\theta}\Delta$, while it is equal to the original $P_h$ inside $B(R_0/2)$. In our applications the angle parameter $\theta\in (0,\pi/4)$ will be assumed small.
Through the twisting $P_h\to P_{h,\theta}$, the continuous spectrum has been tilted from $\IR_+$ to $e^{-2i\theta}\IR_+$, and by doing so has {\em uncovered} the resonances $z_j(h)$ contained in this corresponding sector: these resonances have been turned into eigenvalues, with eigenfunctions $\tilde u_{j}\in L^2$. 
For $h>0$ small enough, the rectangle $R(E,Ch,\gamma h)$ will be contained in the $e^{-2i\theta}$ sector, so we are lead to analyze the (discrete) $L^2$ spectrum of the nonselfadjoint semiclassical operator $P_{h,\theta}$ inside this rectangle.

Let us analyze the twisted Schr\"odinger evolution. We have seen in Section~\ref{s:semiclass} that a wavepacket $u_{\rho_0}$ centered at a phase space point $\rho_0$ is transported by the unitary Schr\"odinger propagator $e^{-itP_h/h}$ along the trajectory $\rho(t)=\varphi^t_p(\rho_0)$. The twisted propagator $\cU^t_\theta = e^{-itP_{h,\theta}/h}$ also transports the wavepacket along the trajectory $(\rho(t))$, but the nonselfadjoint character of $P_{h,\theta}$ will have the effect to modify the norm of the wavepacket:
$$
\frac{d}{dt} \|u(t)\|^2 = \frac{2}{h}\im  \la u(t),P_{h,\theta}u(t)\ra\approx \frac{2\im p_{\theta}(\rho(t))}{h} \|u(t)\|^2\,,
$$
where $p_\theta$ is the principal symbol of $P_{h,\theta}$. When $x(t)$ is outside $B(R_0)$, this symbol reads $p_{\theta}(x,\xi)=e^{-2i\theta}|\xi|^2$, so at the point $\rho(t)$ its imaginary part is $-\sin(2\theta)E<0$. As a result, the norm of $u(t)$ decreases very fast: its norm is reduced to $\cO(h^\infty)$ as soon as $\rho(t)$ exits $B(R_0)$: the twisted propagator is {\em strongly absorbing} outside $B(R_0)$.

\subsection{Resonances vs. classical trapped set}
As explained before, the distribution of resonances in rectangles $R(E,Ch,\gamma h)$ depend crucially on the dynamics of $\varphi^t_p$ on the trapped set $K_E$. Let us explain more precisely how this connection operates, starting with the simple case of a nontrapping dynamics.

\subsubsection{Case of a nontrapping dynamics}\label{s:emptyset}
If $K_E=\emptyset$, any point $\rho_0\in p^{-1}(E)$ will leave $B(R_0)$ within a finite time $T_0$. As a result, a wavepacket $u_{\rho_0}$ microlocalized on $\rho_0$ will be transported by $\cU^t_\theta$ outside of $B(R_0)$, and will be absorbed. Let us now assume that $v_z\in L^2(M)$ satisfies $(P_{h,\theta}-z)v_z=0$, for some $z\in R(E,Ch,\gamma h)$. Elliptic estimates show that $v_z$ can be decomposed as a sum of (normalized) coherent states centered inside a small neighbourhood $U(E)$ of $p^{-1}(E)\cap T^*B(R_0)$:
\bequ\label{e:integral}
v_z = \int_{U(E)}\frac{d\rho}{(2\pi h)^d} \la u_\rho,v_z\ra\, u_\rho + \cO(h^\infty)\,.
\eequ
Let us apply the propagator $\cU^{T_0}_\theta$ to the above equality. On the right hand side each evolved wavepacket 
$\cU^{T_0}_\theta u_\rho = \cO(h^\infty)$ from the above discussion, while on the left hand side we get $\cU^{T_0}_\theta v_z=e^{-iz T_0/h} v_z$. The equality between both sides contradicts our assumption $\im z\geq -\gamma h$. This argument shows that if $K_E=\emptyset$, deeper rectangles $R(E,Ch, \gamma h|\log h|)$ are also empty of resonances \cite{Mart02}.

\subsubsection{Fractal hyperbolic trapped set}
We now consider a nontrivial hyperbolic trapped set $K_E$. In this cases resonances generally exist in $R(E,Ch, \gamma h)$, at least when $C$ and $\gamma$ are large enough. In Section~\ref{s:examples} we have mentioned the case where $K_E$ is composed of a single hyperbolic periodic orbit, for which one can derive explicit asymptotic expressions for the resonances. In case of a more complex, fractal chaotic trapped set, we don't have any explicit expressions at our disposal. Yet, semiclassical methods provide upper bounds for the number of resonances inside $R(E,Ch, \gamma h)$, in terms of the Minkowski dimension of the trapped set $K_E$. 
\begin{thm}[Fractal Weyl upper bound]\label{th1}
Assume the trapped set $K_E$ is a hyperbolic repeller of upper Minkowski dimension $1+2\delta$. Then, for any $C,\gamma>0$, there exits $C_{C,\gamma}>0$ and $h_0$ such that
\bequ\label{e:FWub}
\forall h< h_0,\qquad \# \Res(P_h)\cap R(E,Ch,\gamma h) \leq C_{C,\gamma} h^{-\delta}\,.
\eequ
\end{thm}
The Minkowski dimension is a type of fractal dimension, often called "box dimension". Essentially, it indicates that the volumes of the $\eps$-neighbourhoods of $K_E$ (inside $p^{-1}(E)$) decay as $\eps^{2d-1-(1+2\delta)}$ when $\eps\to 0$. 

The above theorem was first proved in \cite{Sjo90} (for wider rectangles), and then refined by \cite{SjZw07}, both in the case of smooth symbols $p(x,\xi)$. The case of Schottky hyperbolic surfaces was addressed by \cite{Zw99} using semiclassical methods, and generalized to hyperbolic manifolds of higher dimension in \cite{GLZ04} by using transfer operators. The case of scattering by $N\geq 3$ convex obstacles was tackled in \cite{NSZ14}, using quantum monodromy operators (quantizations of Poincar\'e maps). 

The bound \eqref{e:FWub} is called a {\em fractal Weyl upper bound}, by analogy with the selfadjoint semiclassical Weyl's law. Indeed, assume we add to $P_h$ a confining potential $\tilde V(x)$, so that any energy shell $\tilde p^{-1}(E)$ is compact. The spectrum of $\tilde P_h$ is then discrete, and the following semiclassical Weyl's law holds near noncritical energies $E$:
\bequ\label{e:Weyl}
{\rm Spec}(\tilde P_h)\cap [E-Ch,E+Ch] = \frac{1}{(2\pi h)^d}\,{\rm Vol}\big(\tilde p^{-1}([E-Ch,E+Ch])\big) + \cO(h^{-d+1})\,.
\eequ
The volume on the right hand side behaves as $C\mathcal{V}_E h^{-d+1}$ for some $\mathcal{V}_E>0$, while the trapped set $\tilde K_E$ has dimension $1+2(d-1)$, so the power in the above estimate agrees with \eqref{e:FWub}.

\medskip

The result \eqref{e:FWub} and the above selfadjoint Weyl's law differ on several aspects:
\begin{enumerate}
\item \eqref{e:FWub} is an upper bound, not an asymptotics. Numerical studies have suggested that this upper bound should be sharp at the level of the order $h^{-\delta}$, at least if $\gamma$ is large enough. Yet, proved lower bounds  for the counting function are of smaller order $\cO(1)$, similar with the case of a single hyperbolic orbit. A counting function $\asymp h^{-\delta}$ could already be called a {\em fractal Weyl's law}.
\item If a more precise estimate should hold, what could be the optimal constant $C_{C,\gamma}$? How does it depend on the depth $\gamma$? This question is related with the gap question discussed in the next section.
\end{enumerate}
This conjectural fractal Weyl's law has been tested numerically on
various chaotic systems, with variable success:  Schr\"odinger
operator with a smooth potential \cite{Lin02}, hyperbolic surfaces by
\cite{GLZ04} and \cite{Bor14}, discrete time analogues of scattering
systems (quantized open maps) in \cite{NoZw07}, and even
experimentally in the case of the scattering by $N$ disks, see \cite{Kuhl+12}.  

\subsubsection{Sketch of the proof of the Fractal Weyl upper bound}\label{s:FWL-pf}
The spectrum of a nonselfadjoint operator $Q$ is notoriously harder to identify than in the selfadjoint case. To study the spectrum of $Q$ near some value $z_0$, one method is to "hermitize" the operator $Q$, namely study the bottom of the spectrum of the positive operator $(Q-z_0)^*(Q-z_0)$, or equivalently the small {\em singular values} of the operator $Q-z_0$; estimates on the number of singular values will then, through Weyl's inequalities, deliver upper bounds on the number of small eigenvalues of $Q-z_0$. It is much more difficult to obtain lower bounds on the number of eigenvalues: this difficulty explains the large gap between upper and lower bounds.

In our problem, to obtain a sharp upper bound we need to twist again the operator $P_{h,\theta}$, by conjugating it with an operator $G_h$ obtained by quantizing a well-chosen {\em escape function} $g(x,\xi)$:
$$
P_{h,G}\defeq  e^{-G_h} P_{h,\theta}\,e^{G_h}\,.
$$
Through this conjugation, the symbol of the operator can be expanded as 
$$
p_{G}= p_{\theta} - ih \{p_\theta,g\} + smaller\,,
$$
where the Poisson bracket $\{p_\theta,g\}$ represents the time derivative of $g(\rho(t))$. Using the hyperbolicity of the flow, for any $\gamma>0$ it is possible to construct a function $g$ such that  $\{p_\theta,g\}(\rho) \geq 2\gamma h$ as soon as ${\rm dist}(\rho,K_E)\geq h^{1/2}$: this function is called an ``escape function", because it grows along the flow, strictly so outside of the neighbourhood 
$$
K_E(h^{1/2})\defeq \{\rho\in p^{-1}(E); {\rm dist}(\rho,K_E)\leq h^{1/2}\}.
$$ 
As a result, $\im p_G(\rho)\leq - 3/2\gamma h$ for $\rho$ outside $K_E(h^{1/2})$. The above hermitization techniques imply that the eigenstates of $P_{h,G}$ with eigenvalues $z\in R(E,Ch,\gamma h)$ must be concentrated in $K_E(h^{1/2})$. Applying the selfadjoint Weyl's law \eqref{e:Weyl} to this set (thickened to an $h$-energy slab), and expressing its volume in terms of the Minkowski dimension of $K_E$, leads to the bound \eqref{e:FWub}. \qed

\subsubsection{Improved fractal upper bounds on hyperbolic surfaces}
Eventhough the dynamics of $\varphi^t_p$ on $K_E$ is used to construct the escape function, the upper bound \eqref{e:FWub} only depends on the geometry of $K_E$, and not really on the flow $\varphi^t_p$ itself. More recently, finer techniques have been developed in the special case of hyperbolic surfaces, taking into account more efficiently the dynamics on $K_E$ \cite{Nau14,Dya18}. In this case the Minkowski dimension of $K_E$ is given by $1+2\delta$, with $\delta\in (0,1)$ the dimension of the limit set $\Lambda_\Gamma$. The upper bound now has a threshold at the value $\gamma_{th}=\frac{1-\delta}{2}$, which corresponds to the decay rate of a cloud of classical particles. For $\gamma\geq \gamma_{th}$ ("deep resonances") the upper bound remains $\cO(h^{-\delta})$, but for  $\gamma<\gamma_{th}$ (``shallow resonances") the upper bound is of the form $\cO(h^{-\alpha(\gamma)})$, with  $\alpha(\gamma)<\delta$ an explicit function, which decreases when $\gamma\searrow 0$. 
\cite{JaNa12} have actually conjectured that for $\gamma<\gamma_{th}$ and $h$ small enough, the rectangle $R(E,Ch,\gamma h)$ should be empty of resonances. This conjectured gap has not been confirmed numerically.

\section{Dynamical criteria for resonance gaps}

Let us now come to the question of resonance gaps. As explained in the introduction (see \eqref{e:expansion1}), in the case of the wave equation in odd dimension, a global resonance gap ensures that the time evolved wave locally decays at a precise rate. Such a gap therefore reflects the phenomenon of dispersion of the wave, which spreads (leaks) outside any given ball. In the semiclassical setting, we have seen in Section~\ref{s:emptyset} that this leakage is easy to understand if the classical flow is nontrapping: in that case the leakage operates in a finite time $T_0$, following the classical escape of all trajectories.

When there exist trapped trajectories, the explanation of this leakage is more subtle, and requires to take into account the dynamics for long times. In the present situation, this dispersion is induced by a combination of two factors: the hyperbolicity of the classical flow on $K_E$, and Heisenberg's uncertainty principle, which asserts that a quantum state cannot be localized in a phase space ball of radius smaller than $h^{1/2}$.

Our main result, reproduced from \cite{NoZw09}, shows that the rate of this dispersion can be estimated by a certain topological pressure of the flow $\varphi^t_p\restriction_{K_E}$ (see \eqref{e:pressure}), which combines both the unstability of the flow with its complexity.
\begin{thm}[Pressure gap]\label{th2}
Assume that the trapped set $K_E$ is a hyperbolic repeller, and that the topological pressure $\cP(1/2)<0$. Then, for any $\eps>0$, $C>0$, and for $h>0$ small enough, the operator $P_h$ has no resonance in the rectangle $R(E,Ch,(|\cP(1/2)|-\eps) h)$. 
\end{thm}
According to our discussion in Section~\ref{s:chaos}, the pressure $\cP(1/2)$ can take either positive or negative values, respectively in the case of "thick" or "thin" trapped sets. So the condition $\cP(1/2)<0$ characterizes systems with a "thin" enough trapped set. We notice that this bound is sharp in the case $K_E$ consists in a single hyperbolic orbit (Section~\ref{s:single}): in dimension $d=2$, the pressure $\cP(1/2)=-\lambda/2$, which asymptotically corresponds to the first line of resonances.

This pressure bound was proved by \cite{Pa76} in the case of hyperbolic surfaces, by showing that the zeros of the Selberg zeta function satisfy $\re s_j\leq\delta$. In this case, the negativity of the pressure is equivalent with the bound $\delta<1/2$ (see Section~\ref{s:hyperb-surf} for the notations). 

This pressure bound was proved in the case of scattering by $N\geq 3$ disks in $\IR^2$, almost simultaneously and independently by \cite{Ika88} and by \cite{GaRi89} (although the latter article does not satisfy the standards of mathematical rigour, it contains the crucial ideas of the proof, and was the first one to identify the pressure). The method used in \cite{NoZw09}, which we sketch below, relies on similar ideas as these articles, carried out in the general setting of a Schr\"odinger operator $P_h$.

\subsection{Evolution of an individual wavepacket}
Our aim is to show that if $v_z$ is an eigenstate of $P_{h,\theta}$ with eigenvalue $z\approx E$, then $\im z/h \leq \cP(1/2)+\eps$. To do so we will study the propagation of $v_z$ by the Schr\"odinger flow $\cU_{\theta}^t = e^{-itP_{h,\theta}/h}$ for long times (we will need to push the evolution up to {\em logarithmic times} $t\sim C|\log h|$, with $C>0$ independent of $h$). From the decomposition \eqref{e:integral} into wavepackets, we see that it makes sense to study in a first step the evolution of individual wavepackets $u_\rho$, centered at some point $\rho$ in the neighbourhood $U(E)$.

\subsubsection{Hyperbolic dispersion of a wavepacket}\label{s:disp-CS}
Take a wavepacket $u_0$ centered on a point $\rho_0\in K_E$; its semiclassical evolution transports it along $\varphi^t_p(\rho_0)$, but also stretches the wavepacket along the unstable direction $E^u(\rho(t))$, following the linearized evolution $d\varphi^t_p(\rho_0)$.
This spreading can be understood from a simple 1-dimensional toy model, namely the Hamiltonian $q(x,\xi)=\lambda x\xi$, generating the Hamiltonian flow $x(t)=e^{\lambda t} x_0$, $\xi(t)=e^{-\lambda t} \xi_0$, a clearly hyperbolic dynamics.
The quantum evolution is generated by $P_h = \lambda (x \frac{h}{i} \partial_x - ih/2)$; its propagator is a unitary dilation:
\bequ\label{e:linear}
e^{-itP_h/h} u_0(x) = e^{-t\lambda/2} u_0(e^{-t\lambda}x)\,.
\eequ
If we start from a the coherent state $u_0(x)=C_h e^{-\frac{x^2}{2h}}$ centered at the origin, the wavepacket at time $t>0$ will have a horizontal (=unstable) width $e^{t\lambda }h^{1/2}$, while its amplitude will be reduced by a factor $e^{-t\lambda/2}$. The dynamics has dispersed the wavepacket along $E^u$.

\begin{figure}[htbp]
\caption{\label{f:deloc} Left: evolution of a minimal-uncertainty wavepacket: the evolved state stretches exponentially along the unstable directions. By the time $T_E$ the state spreads outside a single cell $V_a$. Right: sketch of the partition $(V_a)$, representing only the elements covering $K_E$. }\begin{center} 
\includegraphics[width=.65\textwidth]{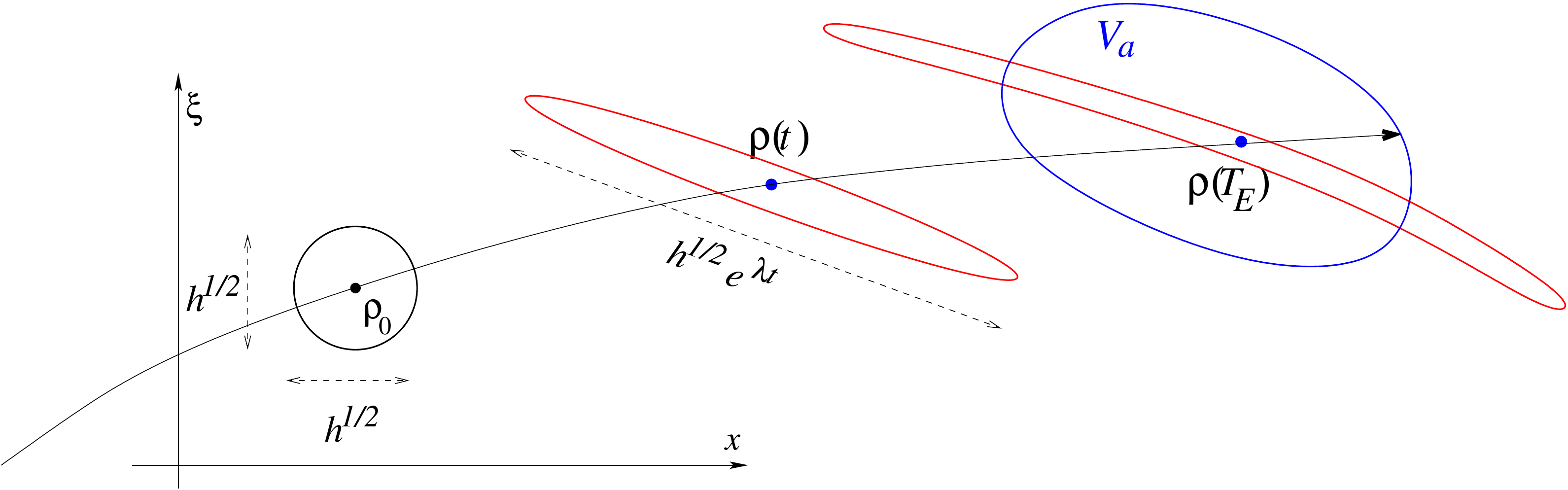}
\hspace{.5cm}
\includegraphics[width=.25\textwidth]{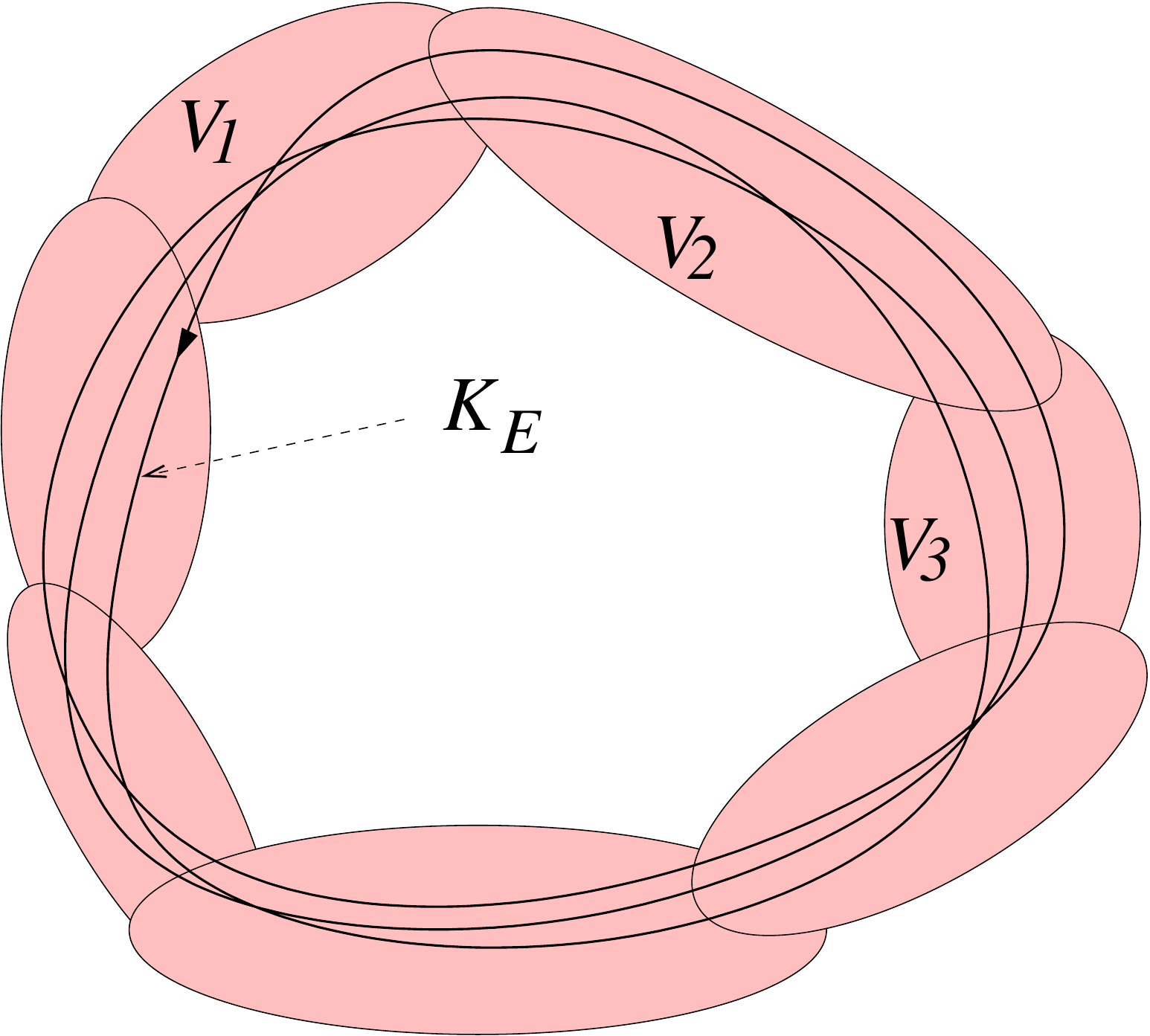}
\end{center}
\end{figure}

Let us come back to our flow $\varphi^t_p$, and assume for simplicity that all the orbits of $K_E$ have the same expansion rate $\lambda>0$, in all unstable directions; this is the case for instance for the geodesic flow in constant curvature $\kappa=-\lambda^2$. In that case, the evolved wavepacket $u(t)$ spreads on a length $\sim e^{t\lambda }h^{1/2}$ along the unstable directions. By the time 
\bequ
T_E=\frac{|\log h|}{2\lambda},\quad\text{which we call the {\em Ehrenfest time}},
\eequ
the wavepacket $u(t)$ spreads on a distance $\sim 1$ along the unstable manifold $W^u(\rho(t))$, it is no more microscopic but becomes macroscopic. Some parts of $u(t)$ are now at finite distance from $K_E$; after a few time steps they will exit the ball $B(R_0)$ and hence be absorbed by the nonunitary propagator (see the left of Fig.~\ref{f:deloc} for a sketch of this evolution).
\subsubsection{Introducing a quantum partition}
In order to precisely estimate the decay of $\|u(t)\|$, one needs to partition the phase space, such as to keep track of the portions of $u(t)$ which exit $B(R_0)$ (and are absorbed), and the ones which stay near $K_E$. One cooks up a finite partition $(V_a)_{a\in A}$ of the phase space $T^*M$ (making it more precise near $K_E$), and quantizes the functions $\bbbone_{V_a}$ to produce a family of microlocal truncations $\Pi_a$, satisfying $\sum_{a\in A} \Pi_a=Id_{L^2}$. The family $(\Pi_a)_{a\in A}$ is called a quantum partition.  

We may insert this quantum partition at each integer step of the evolution: calling $\cU_\theta=e^{-iP_{h,\theta}/h}$, we have for any time $N\in \IN$:
$$
(\cU_\theta)^N = \sum_{\vec a=a_0,\cdots,a_{N}} \cU_{\vec a},\qquad \cU_{\vec a} = \Pi_{a_N}\cU_\theta \cdots \Pi_{a_2} \cU_\theta \Pi_{a_1} \cU_\theta \Pi_{a_0}\,,
$$
where we sum over all possible words $\vec a$ of length $N+1$.
We can control the action of the truncated propagators $\cU_{\vec a}$ on our wavepacket. For times $N<T_E$, the evolved state $u(N)=\cU_\theta^{N} u_0$ is dominated by a single term $\cU_{\vec a}\,u_0$, where the word $\vec a$ is such that each point $\rho(j)\in V_{a_j}$. Around the Ehrenfest time $u(T_E)$ becomes macroscopic, so it is no more concentrated inside a single set $V_a$; the truncations  $\Pi_a$ will cut this state into several pieces, each one carrying a reduced norm. At each following time step,
the evolution $\cU_\theta$ continues to stretch the pieces $\cU_{\vec a}\,u_0$ by a factor $e^\lambda$ along the unstable directions, so several truncations will again act nontrivially. The norms of the pieces $\cU_{\vec a}\,u_0$ can be estimated by the decay of the amplitude of the wavepacket, similarly as in the linear model \eqref{e:linear} (there are now $(d-1)$ unstable directions):
\bequ\label{e:dispersion1}
\| \cU_{\vec a}\, u_0 \|\leq \exp\big(-\frac{\lambda(d-1)}{2} (N-T_E)\big)\,\|u_0\| + \cO(h^\infty),\quad N\geq T_E,\quad \vec a=a_0\cdots a_{N}\,.
\eequ
For most words $\vec a$, this bound is not sharp. For instance, the symbols $a_j$ corresponding to partition elements $V_{a_j}$ outside of $B(R_0)$ indicate that the state is absorbed fast, and lead to $\cO(h^\infty)$ terms. As a result,
for $N>T_E$ the nonnegligible pieces correspond to words $\vec a$ such that almost all the elements $V_{a_j}$ intersect the trapped set. Keeping only those "trapped" words, we obtain
$$
\cU_\theta^N u_0 = \sum_{\vec a\ {\rm trapped}} \cU_{\vec a}\, u_0+{\rm negligible}\,,
$$
with each term bounded as in \eqref{e:dispersion1}. A more careful analysis (involving a "good" choice of partition) shows that for long logarithmic times $N=C|\log h|$, $C\gg 1$, the number of relevant words is bounded above by $\exp(N(H_{top}+\eps))$, where $H_{top}$ is the topological entropy \eqref{e:entropy}, and $\eps>0$ can be made arbitrary small by taking $C$ large enough.

\subsection{Evolving a general state}
Take an eigenstate $v_z$  with eigenvalue $z$ near $E$. Being microlocalized near $p^{-1}(E)$, $v_z$ can be decomposed into wavepackets $u_{\rho}$ as in \eqref{e:integral}. By linearity, we find 
$$
\cU_\theta^N v_z =\frac{1}{(2\pi h)^d}\sum_{\vec a\ {\rm trapped}}\int_{U(E)} d\rho\,  \la u_\rho,v_z\ra\, \cU_{\vec a}\,u_{\rho}\,+\cO(h^\infty)\,,
$$
where each term $\cU_{\vec a}\,u_{\rho}$ is bounded as in \eqref{e:dispersion1}. Applying the triangle inequality, we find
$$
\| \cU_\theta^N v_z \|\leq \frac{{\rm Vol}(U(E))}{(2\pi h)^d} e^{N(H_{top}+\eps)} e^{-\frac{\lambda(d-1)}{2} (N-T_E)}\,.
$$
For a constant expansion rate, $\cP(1/2)= H_{top} -
\frac{\lambda(d-1)}{2}$, so the above bound can be recast as
$h^{-\beta} e^{N(\cP(1/2)+\eps)}$ for some $\beta>0$. Taking $N=C|\log
h|$ with $C$ large enough, we can have $h^{\beta}\leq e^{N\eps}$,
thereby giving a bound $e^{N(\cP(1/2)+2\eps)}$. This bound is
nontrivial if $\cP(1/2)$ is negative. Using the fact that $v_z$ is an
eigenstate of eigenvalue $z\approx E$, we get for such a time $N$:
$$
|e^{N\im z/h} |\leq e^{N(\cP(1/2)+2\eps)} \Longrightarrow \im z/h \leq \cP(1/2)+2\eps\,.
$$\qed

\subsection{Improving the pressure gap}
In the case of a fractal hyperbolic repeller, the pressure bound of Thm~\ref{th2} is believed to be nonoptimal, at least for generic hyperbolic systems. Estimating $\| u(N) \|$ by adding the norms of the terms $\cU_{\vec a}u_\rho$ does not take into account the partial cancellations between these terms. Indeed, when $N=C|\log h|$ with $C\gg 1$, many of those terms are almost proportional to each other, essentially differing by complex valued prefactors. The norm of their sum is hence governed by a sum of many complex factors, which is generally much smaller than the sum of their moduli. 

Such partial cancellations (or ``destructive interferences") are at the heart of Dologopyat's proof of the exponential decay of correlations for Anosov flows \cite{Dol98}, when analyzing the spectrum of a family of transfer operators.
\cite{Nau05} adapted Dolgopyat's method to show an improved high frequency resonance gap for the Laplacian on Schottky hyperbolic surfaces, still working at the level of transfer operators.  By a similar (yet, more involved) method, Petkov and Stoyanov improved the high frequency resonance gap for scattering by convex obstacles on $\IR^d$; these authors managed to establish a semiclassical connection between the quantum propagator and a transfer operator, thereby applying Dolgopyat's method to the former. All the above works improve the pressure bound by some small, not very explicit $\eps_1>0$. 

In the case of hyperbolic surfaces, a recent breakthrough was obtained by Dyatlov and his collaborators. \cite{DZ16} showed that a nontrivial gap for a hyperbolic surface with parameter $\delta\in (0,1)$ results from a {\em fractal uncertainty principle} (FUP), a new type of estimate in 1-dimensional harmonic analysis. This FUP states that if $K\subset [0,1]$ is a Cantor set of dimension $\delta$ and $K(h)$ its $h$-neighbourhood, then there exists $\beta>0$ such that 
$$
\| \bbbone_{K(h)} \cF_h \bbbone_{K(h)} \|_{L^2\to L^2}\leq C h^\beta\,,
$$
where $\cF_h$ is the semiclassical Fourier transform. This estimate shows that a function $u\in L^2(\IR)$ and its semiclassical Fourier transform cannot be both concentrated on $K(h)$. This FUP obviously holds when $\delta<1/2$, giving back the pressure bound. In a ground-breaking work \cite{BoDy16} managed to prove this FUP in the full range $\delta\in (0,1)$, thereby showing a resonance gap on any Schottky hyperbolic surface. The improved gap is not very explicit, it is much smaller than the gap $\frac{1-\delta}{2}$ conjectured by Jakobson-Naud.

Although the methods of \cite{DZ16} strongly rely on the constant negative curvature, it seems plausible to prove a resonance gap for any hyperbolic repeller in two space dimensions. On the other hand, the extension of an FUP to higher dimensional systems is at present rather unclear, partly due to the more complicated structure of the trapped sets.

\section{Normally hyperbolic trapped set}\label{s:NHIM}
\begin{figure}[htbp]
\caption{\label{f:NHIM} Sketch of a normally hyperbolic trapped set $K$. }
\begin{center} 
\includegraphics[width=.5\textwidth]{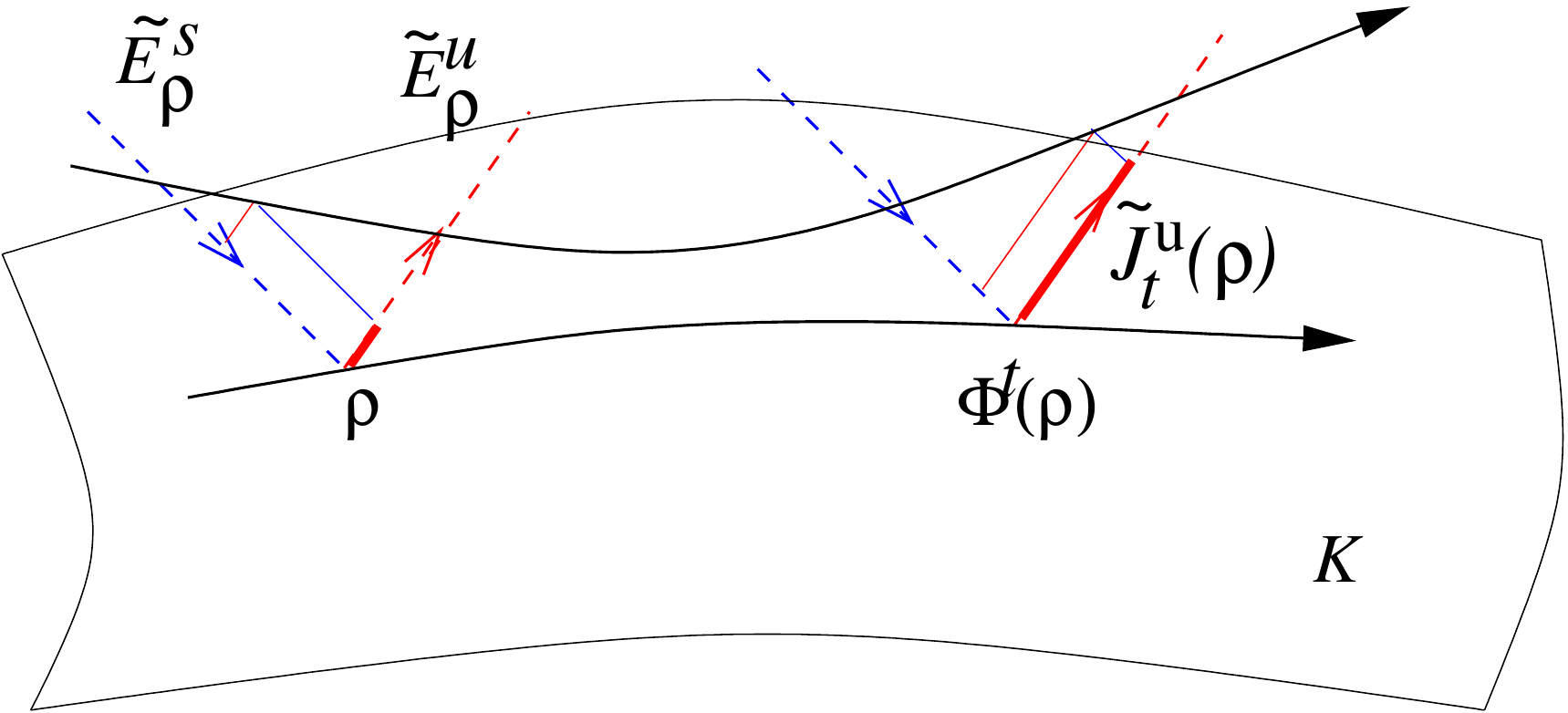}
\end{center}
\end{figure}

In this last section, we focus on a different type of trapped set. We assume that for some energy window $[E_1,E_2]$, the trapped set $K=K_{[E_1,E_2]}=\cup_{E\in [E_1,E_2]}K_E$ is a smooth, normally hyperbolic, symplectic submanifold of the energy slab $p^{-1}([E_1,E_2])$. What does this all mean?
If $K$ is a symplectic submanifold of $T^*M$, at each point $\rho\in K$ the tangent space $T_\rho(T^*M)$ splits into $T_\rho K \oplus (T_\rho K)^\perp$, where both are symplectic subspaces. Normal hyperbolicity means that the flow $\varphi^t_p$ is hyperbolic transversely to $K$: the transverse subspace 
$(T_\rho K)^\perp=\tilde E^s(\rho)\oplus \tilde E^u(\rho)$, such that $d\varphi^t_p\restriction_{TK^\perp}$ contracts exponentially along $\tilde E^s(\rho)$, and expands along  $\tilde E^u(\rho)$ (see Fig.~\ref{f:NHIM}). We denote by $\tilde J^u_t(\rho)= |\det (d\varphi^t_p\restriction_{\tilde E^u_\rho})|$ the normal unstable Jacobian.

\subsection{Examples of normally hyperbolic trapped sets}
\subsubsection{Examples in chemistry and general relativity}
This dynamical situation may occur in quantum chemistry, when modeling certain reaction dynamics. The reactants and products of the chemical reaction are two parts of phase space, connected by a hyperbolic ``saddle" along two conjugate coordinates $(x_1,\xi_1)$, similar with the linear dynamics of Section~\ref{s:disp-CS}, while the evolution of the other coordinates remains bounded \cite{Go+10}. The trapped set $K_{[E_1,E_2]}$ is then a bounded piece of the space $\{x_1=\xi_1=0\}$.

This dynamical situation also occurs in general relativity, namely when describing timelike trajectories in the Kerr or Kerr-de Sitter black holes \cite{WuZw11,Dy12}. The trapped set is a normally hyperbolic manifold diffeomorphic to $T^*S^2$. In this situation resonances are replaced by quasinormal modes, obtained by solving a generalized spectral problem $P(z)u=0$. Yet, the semiclassical methods sketched below can be easily adapted to this context.

\subsubsection{From classical to quantum resonances}\label{s:FS}
An original application of this dynamical assumption concerns the study of contact Anosov flows. A flow $\phi^t$ defined on a compact manifold $M$ is said to be Anosov if at any point $x\in M$, the tangent space $T_xM$ splits into $\IR \Xi(x)\oplus E^u(x)\oplus E^s(x)$, where $\Xi(x)$ is the vector generating the flow, while $E^s(x)$, $E^u(x)$ are the stable/unstable subspaces, satisfying the properties \eqref{e:EuEs}. The assumption that $\phi^t$ preserves a contact 1-form $\alpha$, implies that the subspace $E^u(x)\oplus E^s(x)$, which forms the kernel of $d\alpha(x)$, depends smoothly on $x$. 

The long time properties of such a flow are governed by a set of so-called Ruelle-Pollicott (RP) resonances $\{\lambda_k \subset \IC_-\}$, which share many properties with the quantum resonances we have studied so far. Considering two test functions $u,v\in C^\infty(M)$, their {\em correlation function}
$C_{v,u}(t)\defeq \int_M dx\, v(x) u(\phi^t(x)) - \int dx\, v(x) \int dx\, u(x) $ can be expanded in terms of these RP resonances:
\bequ\label{e:correl}
C_{v,u}(t)= \sum_{\im \lambda_k\geq -\gamma} e^{-i\lambda_k t} \la v,\Pi_{\lambda_k}u\ra + \cO_{u,v}(e^{-\gamma t})\,,
\eequ
Hence, if the RP resonances $\lambda_k$ satisfy a uniform gap, the correlation decays exponentially (one speaks of {\em exponential mixing}). Such a resonance gap has been first proved by \cite{Dol98} and \cite{Liv04}, while \cite{Tsu10} proved an explicit bound for the high frequency gap. 

Comparing \eqref{e:correl} with \eqref{e:expansion1}, \cite{FS10} had the idea to interpret the RP resonances (or rather $z_k=h\lambda_k$) as the ``quantum resonances" of the ``quantum Hamiltonian" $P_h = -ih\Xi$. Notice that $e^{-itP_h/h}u(x)=u(\phi^{-t}(x))$. What do we gain from this interpretation?
The principal symbol of $P_h$, $p(x,\xi)=\xi(\Xi(x))$, generates on $T^*M$ the symplectic lift of $\phi^t$ :
$\varphi^t_p(x,\xi) = (\phi^t(x), ^T\!\!d\phi^t(x)^{-1}\xi)$. As opposed to the scattering situation, each energy shell $p^{-1}(E)$ goes to infinity along the fibers of $T^*M$. Hence, for any energy $E\in\IR$, the trapped set $K_E$ is given by the points $\rho=(x,\xi)\in p^{-1}(E)$ such that $^T\!d\phi^t(x)^{-1}\xi$ remains bounded when $t\to\pm\infty$. From the hyperbolicity structure, this is possible only if $\xi=E\alpha_x$. Hence, $K_E=\{(x,\xi=E\alpha_x),\,x\in M\}$, a smooth submanifold of $p^{-1}(E)$. It is easy to check that $K=\cup_E K_E$ is symplectic, and normally hyperbolic (the subspaces $\tilde E^{s/u}$ are lifts of the subspaces $E^{s/u}$ of $TM$). The resonances of the quantum Hamiltonian $P_h$ can thus be connected with the properties of this trapped set.

The main difficulty when analyzing this classical dynamical problem as a ``quantum scattering" one \cite{FS10}, is to twist the selfadjoint operator $P_h$, such as to transform the resonances into eigenvalues. This was done by constructing
spaces of anisotropic distributions $\cH^m\subset \cD'(M)$, such that $P_h:\cH^m\to \cH^m$ has discrete spectrum in $\{\im z\geq -mh\}$, made of "uncovered" Ruelle-Pollicott resonances. We will not detail this construction, which can also be presented as a twist of the operator $P_h$ into a nonselfadjoint operator $P_{h,m}$ on $L^2(M)$.

\subsection{An explicit resonance gap for normal hyperbolic trapped sets}
Let us come back to our general setting, and start again to propagate minimum-uncertainty wavepackets $u_\rho$ centered on a point $\rho\in K$.  Due to the normal hyperbolicity, the state $e^{-itP_h/h}u_\rho$ spreads exponentially fast along the transverse unstable direction $\tilde E^u$. Similarly as what we did in Section~\ref{s:FWL-pf}, one can twist the operator $P_h$ by a microlocal weight $G_h$, such that the twisted operator $P_{h,G}$ is absorbing outside the neighbourhood $K(Ch^{1/2})$. After a few time steps, the evolved wavepacket will leak outside of this neighbourhood, and be partially absorbed: their norms will decay at the rate
$$
\|e^{-it P_{h,G}/h}u_\rho\|\leq C\,\tilde J^u_t(\rho)^{-1/2}\,,\qquad t>0.
$$
If we call $\tilde \Lambda_{\min} = \liminf_{t\to \infty}\frac{1}{t}\inf_{\rho\in K} \log \tilde J^u_t(\rho)$ the minimal growth rate of the transverse unstable Jacobian, for any $\eps>0$ and $t>t_\eps$ large enough, the above right hand sides are bounded by $e^{-t(\tilde\Lambda_{\min}/2-\eps)}$. With more work, one can show that this uniform decay of our individual wavepackets induces the same decay of any state microlocalized on $K$, in particular of any eigenstate $v_z$ of $P_{h,G}$. One then obtains the following gap estimate for the eigenvalues of $P_{h,G}$, or equivalently the resonances of $P_h$ \cite{NoZw15}:
\begin{thm}[Resonance gap, normally hyperbolic trapped set]
Assume the trapped set $K=K_{[E-c,E+c]}$ is normally hyperbolic, with minimal transverse growth rate $\tilde\Lambda_{\min}$. Then, for any $\eps>0$ and $h>0$ small enough, the rectangle $R(E,c, (\tilde\Lambda_{\min}/2-\eps)h)$ contains no resonance.
\end{thm}
Like in the case of Thm~\ref{th2} and its improvements, we also obtain a bound for the truncated resolvent operator inside the rectangle, of the form $\|\chi(P_h-z)^{-1}\chi\|\leq h^{-\beta}$, $\chi\in C^\infty_c(M)$. 
When applying this result to the situation of Section~\ref{s:FS} (mixing of contact Anosov flows), we exactly recover Tsujii's gap for the high frequency RP resonances. 
\medskip

In two of the settings presented above (the resonances of Kerr-de Sitter spacetimes \cite{Dy16}, respectively the Ruelle-Pollicott for contact Anosov flows \cite{FTs13}, the spectrum of resonances has been shown to enjoy a richer structure, provided certain bunching conditions on the rates of expansion are satisfied. Namely, beyond the first gap stated in the above theorem, resonances are gathered in a (usually finite) sequence of parallel strips, separated by secundary resonance free strips. The widths of the strips are expressed in terms of maximal and minimal expansion rates similar with $\Lambda_{\min}$. Besides, the number of resonances along each of the strips satisfies a Weyl's law, corresponding to the volume of the $h^{1/2}$ neighbourhood of $K$. 


\subsection*{Acknowledgments}
The author has benefitted from invaluable interactions with many colleagues, in particular M.Zworski who introduced him to the topic of chaotic scattering, as well as V.Baladi, S.Dyatlov, F.Faure, F.Naud, J.Sj\"ostrand. In the past few years he has been partially supported by the grant Gerasic-ANR-13-BS01-0007-02 awarded by the Agence Nationale de la Recherche. 



\end{document}